\documentclass[12pt]{article}\usepackage{RhoFactorize}

\begin{document}

\title{Factorization of \\
Quantum Density Matrices\\
According to \\
Bayesian and Markov Networks}

\author{Robert R. Tucci\\
        P.O. Box 226\\
        Bedford,  MA   01730\\
        tucci@ar-tiste.com}

\date{ \today}

\maketitle

\vskip2cm
\section*{Abstract}
We show that any quantum density matrix
can be represented by a Bayesian network (a directed
acyclic graph), and also  by a Markov network (an
undirected graph).
We show that any Bayesian
or Markov net
 that represents a density matrix,
is logically equivalent to
a set of conditional independencies
(symmetries) satisfied by the density matrix.
We show that the d-separation theorems
of classical Bayesian and Markov networks
generalize in a simple and natural way to
quantum physics. The quantum
d-separation  theorems
are shown to be closely connected
to quantum entanglement.
We show that the graphical rules for d-separation
can be used to
detect
pairs of nodes (or of node sets) in a graph that are unentangled.
CMI entanglement (a.k.a. squashed entanglement),
a measure of entanglement originally discovered
by analyzing Bayesian networks,
is an important part of the theory of this
paper.

\newpage
\section{Introduction}
A Bayesian network is a directed
graph; that is, a set of
nodes
with arrows connecting some
pairs of these nodes. Each
node is assigned a transition matrix.
For a classical Bayesian net,
each transition matrix is real,
and the product the
transition matrices for all the
nodes gives a joint probability
distribution for the states of all the nodes.
For a quantum Bayesian net,
each transition matrix is complex,
and the product of the
transition matrices  gives a joint probability
amplitude instead.

A Markov network is an undirected graph; that is,
a set of nodes with
undirected links
connecting some pairs of these nodes.
Each super-clique (maximal
fully-connected subgraph) of the graph
 is assigned an affinity.
 For a classical Markov net,
each affinity is real,
and their product gives a joint probability
distribution for the states of all the nodes.
For a quantum Markov net,
each affinity is complex,
and their product gives a joint probability
amplitude instead.

Bayesian and Markov networks
will be defined
more precisely later on
in this paper.

The literature on classical Bayesian nets
is vast. Some textbooks
that were invaluable in writing this paper
are Refs.\cite{Kol},\cite{Lau}.
Classical Bayesian nets were invented
by geneticist Sewall Wright\cite{Sew}
 in the early 1930's.
The
theory of Bayesian nets was
 extended substantially by
Judea Pearl\cite{Pea-wiki}\cite{Pea1}\cite{Pea2}
 and collaborators in the late 1980's.
 They gave
us the theory that culminates in
the d-separation rules.
See Scheines\cite{hist} for
 a  more complete review of
the history of d-separation.
Nowadays, classical Bayesian nets
are used widely in Data mining, AI, etc.

There exist only a small
number of papers on quantum Bayesian nets.
The first paper\cite{Tucci-bnet1}
on the subject
appears to be mine.
Since then,
I have written several papers
applying quantum Bayesian nets to
quantum information theory\cite{Tucci-meta-rho1}
and
quantum computing\cite{Tucci-bnet-qcomp}. I have
also written a Mac application
called Quantum Fog\cite{artiste}
(freeware but patented) that implements
the ideas behind quantum Bayesian networks.
Laskey has also written some papers\cite{Kat}
about quantum Bayesian nets.

It's known that any {\it probability distribution}
 can
be represented by a Bayesian net, and also by a
Markov net.
It's known that any Bayesian or Markov net
that represents a {\it probability distribution},
is logically equivalent to
a set of conditional independencies
satisfied by the {\it probability distribution}.

In this paper, we show that the last paragraph is true
if we replace {\it probability distribution} by
{\it density matrix}.

We also show that the d-separation theorems
of classical Bayesian and Markov networks
generalize in a simple and natural way to
quantum physics. The quantum
d-separation  theorems
are shown to be closely connected
to quantum entanglement.
We show that the graphical rules for d-separation
can be used to
detect
pairs of nodes (or of node sets) in a graph that are unentangled.
CMI entanglement (a.k.a. squashed entanglement)\cite{wiki-cmi-ent},
a measure of entanglement originally discovered
by analyzing Bayesian networks,
is an important part of the theory of this
paper.

This paper is fairly self-contained;
readers
previously acquainted with
quantum physics but not
with classical Bayesian nets
should have no trouble
following this paper. Results about
classical Bayesian nets are derived in
parallel with those about their quantum
brethren. The paper has pretensions of
being pedagogical.

\section{Notation and Other Preliminaries}
In this section, we define
some notation, and
review various
prerequisite ideas
that will be used in the rest of the paper.

\subsection{General Notation}

As usual, $\ZZ,\RR, \CC$
will denote the integers, real numbers,
and complex numbers, respectively.
Let $Bool=\{0,1\}$, $0=false$ and $1=true$.
For $a,b\in\ZZ$ such that $a\leq b$,
let
 $Z_{a,b}=\{a,a+1,\ldots, b\}$.

For any set $J$, let $|J|$
denote the number of elements in $J$.

For any set $J$, its power-set
is defined as $\{J':J'\subset J\}$. This set
 includes the empty set $\emptyset$
and the full set $J$. The power-set of
$J$ is often denoted by $2^J$ because
$|2^J| = 2^{|J|}$.

Let
$\delta^x_{y}=\delta(x,y)$
denote the Kronecker delta function;
it equals 1 if $x=y$ and 0 if $x\neq y$.

For any matrix $M\in \CC^{p\times q}$,
$M^*$ will denote its complex conjugate,
$M^T$ its transpose, and $M^\dagger = M^{*T}$
its Hermitian conjugate.
Let $\diag(x_1,x_2,\ldots,x_r)$
denote a diagonal matrix with
diagonal entries $x_1,x_2,\ldots,x_r$.

For any $z\in \CC$,
$phase(z)$ will denote its phase.
If $r,\theta\in \RR$, $phase(re^{i\theta})=\theta
+2\pi \ZZ$.

For any expression $f(x)$,
we will sometimes abbreviate

\beq
\frac{f(x)}{\sum_x f(x)}=
\frac{f(x)}{\sum_x numerator}
\;.
\eeq
The abbreviation
with the word ``numerator" is especially
helpful when $f(x)$ is a
long expression, and we
want to write it
only once instead of twice.

For $f_1,f_1\in \CC$, let

\beq
\biprod{f_1}{f_2} = f_1 \times f_2
\;.
\eeq
This notation saves horizontal space: it
allows us to
indicate the product of two numbers
with the numbers
written in a column instead of a
row.

Given expressions A,B,X,Y, we will often say
things like ``A (ditto, X) is B (ditto, Y)";
by this, we will mean that ``A is B" and ``X is Y".

\subsection{Classical Probability Theory \\
and Quantum Physics Preliminaries}

Random variables will be denoted
by underlined letters; e.g.,
$\rva$.
The set of values (states) that
$\rva$ can assume will be denoted
by $St_\rva$. Let $N_\rva=|St_\rva|$.
\footnote{We will use random variables
in both classical and quantum physics.
Normally, random variables are
defined only in classical physics, where they are
defined to be functions from an outcome space to a
range of values.
For technical simplicity, here we
define a random variable $\rva$, in both
classical and quantum physics,
to be merely the label of a
 node in a graph, or an n-tuple
 $\rvxvec{K}$ of such labels.}
The probability that
$\rva=a$ will be denoted by $P(\rva=a)$
or $P_\rva(a)$, or simply by $P(a)$
if the latter will not lead to confusion
in the context it is being used.
We will use $pd(St_\rva)$ to denote
the set of all probability distributions
with domain $St_\rva$.

In this paper, we consider
networks with $N$ nodes.
Each node is labelled by a random variable
$\rvx_j$,
where $j\in \zn$.
For any $J\subset \zn$,
the ordered set of random
variables $\rvx_j$ $\forall j\in J$
(ordered so that the integer indices $j$
 increase from left to right)
will be denoted by
$(\rvxdot)_J$ or $\rvxvec{J}$.
For example, $(\rvxdot)_{\{2,4\}}=
\rvxvec{\{2,4\}}=(\rvx_2,\rvx_4)$.
We will often call the values that
$\rvxvec{J}$
can assume $(x.)_J$ or $\xvec{J}$.
For example, $(x.)_{\{2,4\}}=
\xvec{\{2,4\}}=(x_2,x_4)$.
We will often abbreviate
$(\rvxdot)_\zn$ or $\rvxvec{\zn}$
by just
$(\rvxdot)$ or $\rvxdot$\;\;.
We will often call the values that
$\rvxdot$
can assume $(x.)$ or $x.$\;\;.

In this paper, we will
often divide by probabilities
without specifying that
they should be non-zero. Most of
the time, this cavalier attitude
will not get us into trouble.
That's because one can always replace all vanishing
probabilities by a positive infinitesimal
$\epsilon$. Our results
can then be expressed as a power series in
$\epsilon$.
As long as our
inferences depend only
on terms that are zeroth order in
$\epsilon$,
our inferences will be well-defined
and unique as $\epsilon$ tends to 0.
There are, however,
situations
when dividing by a probability
can be fatal. Such situations
ultimately boil down to trying to
infer something from terms that are
first order in $\epsilon$; for example,
when we  erroneously conclude
that  $A\epsilon=0$
implies $A=0$.
In the future,
we will divide by
probabilities without
assuming that they should
be non-zero, except in those
cases when doing so is being
used to infer something
that becomes false when
$\epsilon\rarrow 0$.

In quantum physics, $\rva$ has a {\it fixed,
orthonormal} basis
$\{\ket{a}:a\in St_\rva\}$
associated with it.
The vector space spanned
by this basis will be denoted by $\calh_\rva$.
In quantum physics, instead of
probabilities $P(\rva=a)$,
we use ``probability amplitudes"
(or just ``amplitudes" for short)
$A(\rva=a)$ (also denoted by $A_\rva(a)$
or $A(a)$).
Whereas $P\geq 0$ and
$\sum_a P(a)=1$,
$\sum_a |A|^2(a)=1$.
Besides probability amplitudes,
we also use density matrices.
A density matrix
$\rho_\rva$ is a Hermitian,
non-negative, unit trace
operator
acting on $\calh_\rva$.
We will use $dm(\calh_\rva)$ to
denote the set of all
density matrices acting on $\calh_\rva$.

If $\rho_\rvx \in dm(\calh_\rvx)$,
$\rho_{\rvx,\rva}\in dm(\calh_{\rvx,\rva})$,
and $\rho_\rvx = \tr_\rva (\rho_{\rvx,\rva})$,
we will say that $\rho_\rvx$ is a
{\bf partial trace of}
$\rho_{\rvx,\rva}$, and
$\rho_{\rvx,\rva}$ is a {\bf traced dm-extension of}
$\rho_\rvx$.
Given a density matrix
$\rho_{\rvx_1,\rvx_2,\rvx_3,\ldots}\in
dm(\calh_{\rvx_1,\rvx_2,\rvx_3,\ldots})$,
its partial traces will be denoted
by omitting its subscripts for
the random variables that
have been traced over. For example,
$\rho_{\rvx_2} = tr_{\rvx_1,\rvx_3}
\rho_{\rvx_1,\rvx_2,\rvx_3}$.

We will sometimes abbreviate $\ket{a}\bra{a}$
by $\proj(\ket{a})$.
This abbreviation is
especially convenient when the
label $a$ is a long expression,
for then we only have to write
$a$ once instead of twice.

\subsection{Graph Theory Preliminaries}
Next, we review some basic
definitions from Graph Theory.

A {\bf graph} $G$ is pair $(V,E)$,
where $V$ is a set of {\bf nodes (vertices)} ,
and  $E$ is a set of
{\bf connections (edges)} between some pairs of
these nodes. (No self-connections allowed).
A {\bf subgraph} $G'=(V',E')$ of a graph $G=(V,E)$ is a
graph such that $V'\subset V$,
and $E'$ is defined as the subset of $E$
that survives after we erase from $E$ all edges
that mention a node in $V-V'$.

We will abbreviate {\bf Directed Acyclic Graph}
by DAG.
A DAG is a
graph with arrows as its edges,
and without any cycles.
A {\bf cycle} is a finite sequence of arrows
that one can follow, in the direction of
the arrows, and come back to where one started.
The set of all possible DAGs with node labels
$\rvxdot$ will be denoted by $DAG(\rvxdot)$.

We will abbreviate {\bf Undirected Graph} by UG.
An UG is a graph with
(undirected) links as its edges.
The set of all possible UGs with node labels
$\rvxdot$ will be denoted by $UG(\rvxdot)$.

One can also define hybrid graphs
that contain both
arrows and undirected links\cite{Lau}\cite{Kol},
but we won't consider them in this paper.

\begin{figure}[h]
    \begin{center}
    \epsfig{file=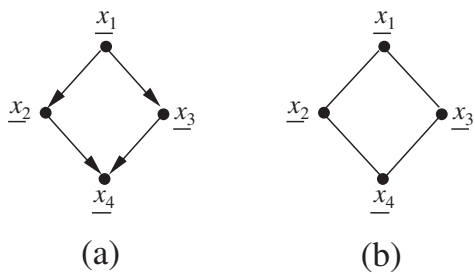, height=1.5in}
    \caption{(a)An example of a Bayesian net.
    (b)An example of a Markov net.}
    \label{fig-diamond-graphs}
    \end{center}
\end{figure}

Consider a DAG whose nodes are labelled by
$\rvxdot$\;\;. Any node $\rvx_j$
has parent nodes (those with arrows
pointing from them to $\rvx_j$)
and children nodes (those with arrows pointing
from $\rvx_j$ to them).
$pa(j), ch(j)\subset \zn$ are defined
as the sets of integer indices
of the {\bf parent and children nodes}
of $\rvx_j$.
For example, in Fig.\ref{fig-diamond-graphs}(a),
$pa(4)=\{2,3\}$ and
$ch(1)=\{2,3\}$.
$an(j), de(j)\subset \zn$
are defined
as the sets of integer indices
of the {\bf ancestor and descendant nodes}
of $\rvx_j$.
That is,
$an(j)=pa(j)\cup pa^2(j)\cup pa^3(j)\cup\ldots$.
By this we mean that $an(j)$
is obtained by
taking the union of the integer indices of the parents of
$\rvx_j$, and of the parents of the parents of
$\rvx_j$, and of the parents of the parents
of the parents of $\rvx_j$,
and so on.
Likewise,
$de(j)=ch(j)\cup ch^2(j)\cup ch^3(j)\cup\ldots$.
The set of integer indices of the
{\bf non-descendants} of
$\rvx_j$ will be denoted by
$\neg de(j)=\zn-de(j)-\{j\}$.
The set of integer indices of the
{\bf non-ancestors} of
$\rvx_j$ will be denoted by
$\neg an(j)=\zn-an(j)-\{j\}$.
Let $\overline{s}(j)=s(j)\cup \{j\}$
for $s\in\{ pa, ch, an, de, \neg de, \neg an\}$.
In other words, we will use an overline over a set
$s(j)$ that does not include $j$
to denote the ``closure" set obtained by adding
$j$ to $s(j)$.

Next consider an UG whose nodes are labelled by
$\rvxdot$\;\;. Any node $\rvx_j$
has {\bf neighbor nodes} (those with links
between $\rvx_j$ and them).
$ne(j)$ is
defined as the set of integer indices of
the neighbor nodes of $\rvx_j$.
For example, in Fig.\ref{fig-diamond-graphs}(b),
$ne(2)=\{1,4\}$.
We will also use
$\overline{ne}(j)=ne(j)\cup \{j\}$.

For
either a DAG or an UG, a {\bf path
from node $\rvx$ to node $\rvy$}
is
a finite sequence of nodes,
starting with $\rvx$ and ending with $\rvy$,
such that adjacent nodes
 in the sequence
are connected. Note
that for a DAG, the arrows in a path
need not all be oriented in the same sense.
If they are, we call the path a
{\bf directed path}.

In a DAG, a path from $\rvx$ to
$\rvy$ can have 3 (mutually exclusive
and exhaustive) types of
nodes. A {\bf serial node} $\rva$
equals one of the endpoints
($\rvx$ and  $\rvy$), or else, it
is connected to its path neighbors
in this
\beq\rarrow(\rva)\rarrow\eeq
or  this
\beq\larrow(\rva)\larrow\eeq
manner.
A
{\bf divergence node} $\rva$
is connected to its path neighbors
in this
\beq\larrow(\rva)\rarrow\eeq
manner.
A
{\bf convergence (a.k.a. collider) node} $\rva$
is connected to its path neighbors
in this
\beq\rarrow(\rva)\larrow\eeq
manner.

A DAG (ditto, an UG) is {\bf fully connected}
if it is impossible to add
any more legal arrows (ditto, links) to it.
A fully connected subgraph (of either a DAG or an UG)
is called a {\bf clique}.
A clique for which there is no
larger clique that contains it, is
called a {\bf super-clique}. For any graph $G$,
we define
$super-cliques(G)$ (a subset of $2^\zn$)
to be the set of the super-cliques of $G$.
For example, $super-cliques(G)$
for both graphs in Fig.\ref{fig-diamond-graphs}
is
$\{\{1,2\}, \{1,3\},\{2,4\},\{3,4\}\}$.

A {\bf classical Bayesian network} is a DAG
with labelled nodes (let
$(\rvxdot)_{\zn}$ be the labels), together with
a transition matrix $P(x_j|\xvec{pa(j)})$
associated with each node $\rvx_j$ of the graph.
The quantities $P(x_j|\xvec{pa(j)})$
are probabilities; they
are non-negative and satisfy
$\sum_j P(x_j|\xvec{pa(j)})=1$.
The probability of the whole net is defined
as the product of the probabilities of the nodes.

A {\bf quantum Bayesian network} is a DAG
with labelled nodes (let
$(\rvxdot)_{\zn}$ be the labels), together with
a transition matrix $A(x_j|\xvec{pa(j)})$
associated with each node $\rvx_j$ of the graph.
The quantities $A(x_j|\xvec{pa(j)})$
are probability amplitudes; they
satisfy
$\sum_j |A|^2(x_j|\xvec{pa(j)})=1$.
The probability amplitude of the whole net is defined
as the product of the amplitudes of the nodes.
For example, for the quantum
Bayesian net of Fig.\ref{fig-diamond-graphs}(a),
one has

\beq
A(x_1,x_2,x_3,x_4) =
A(x_4|x_2,x_3)A(x_3|x_1)A(x_2|x_1)A(x_1)
\;,
\eeq
where $x_j\in St_{\rvx_j}$ for $j=1,2,3,4$.

A {\bf classical (ditto, quantum) Markov network} is an UG
with labelled nodes (let
$(\rvxdot)_{\zn}$ be the labels), together with
an affinity $\phi(\xvec{K})$
(ditto, $\alpha(\xvec{K})$)
associated with each super-clique
$K$ of the graph.
The probability (ditto, probability
amplitude) of the whole net is defined
as the normalized product of the affinities of the
super-cliques of $G$. For example,
for the quantum Markov net of
Fig.\ref{fig-diamond-graphs}(b),
one has

\beq
A(x_1,x_2,x_3,x_4) =
\frac{\alpha(x_4,x_3)\alpha(x_4,x_2)\alpha(x_3,x_1)\alpha(x_2,x_1)}
{\sqrt{\sum_{x_1,x_2,x_3,x_4}\; |numerator|^2}}
\;,
\eeq
where $x_j\in St_{\rvx_j}$ for $j=1,2,3,4$.

We will sometimes use
$\tilde{G}$ to denote a
Bayesian (ditto, Markov) network
associated with a DAG (ditto, UG)  $G$.

\subsection{Information Theory Preliminaries}

Next, we review some basic
definitions from Information Theory\cite{Cover}.

First consider classical physics.
For any $P\in pd(St_\rvx)$,
the {\bf entropy} (a measure of the
variance of $P$) is defined by

\beq
H(\rvx) = -\sum_x P(x)\ln P(x)
\;.
\eeq
Sometimes the entropy is denoted
instead by $H(P_\rvx)$.
CMI (usually pronounced ``see-me")
stands for ``Conditional Mutual Information".
For $P\in pd(St_{\rvx,\rvy,\rvz})$,
the {\bf CMI} (a measure of conditional information
transmission) is defined by

\beq
H(\rvx:\rvy|\rve)= \sum_{x,y,e}
P(x,y,e)\ln \frac{P(x,y|e)}{P(x|e)P(y|e)}
\;.
\eeq
In general, $H(\rvx:\rvy|\rve)\geq 0$.
When $N_\rve=1$, CMI degenerates into
the {\bf mutual information} $H(\rvx:\rvy)$.
Note that

\bsub
\beqa
H(\rvx:\rvy|\rve)&=&
\sum_{x,y,e}
P(x,y,e)\ln \frac{P(x,y,e)P(e)}{P(x,e)P(y,e)}\\
&=& H(\rvx,\rve) + H(\rvy,\rve)
- H(\rvx,\rvy,\rve) - H(\rve)
\;.
\label{eq-h-cmi-expanded}
\eeqa
\esub
Classical CMI satisfies the chain rule

\beq
H(\rvx:\rvy_1,\rvy_2|\rve)=
H(\rvx:\rvy_1|\rvy_2,\rve)+
H(\rvx:\rvy_2|\rve)
\;.
\eeq

Now consider  quantum physics.
For $\rho_\rvx \in dm(\calh_\rvx)$,
the  {\bf entropy}
is defined by

\beq
S(\rvx) = -\tr_x (\rho_\rvx \ln \rho_\rvx)
\;.
\eeq
Sometimes the entropy is denoted
instead by
$S(\rho_\rvx)$
or by
$S_{\rho}(\rvx)$,
where $\rho$ is a traced dm-extension of $\rho_\rvx$.
For $\rho_{\rvx,\rvy,\rve}\in dm(\calh_{\rvx,\rvy,\rve})$,
the {\bf CMI} is defined by
analogy to Eq.(\ref{eq-h-cmi-expanded}):

\beq
S(\rvx:\rvy|\rve)=
S(\rho_{\rvx,\rve}) + S(\rho_{\rvy,\rve})
- S(\rho_{\rvx,\rvy,\rve}) - S(\rho_{\rve})
\;.
\eeq
In general, $S(\rvx:\rvy|\rve)\geq 0$ (this is
known as the Strong Subadditivity of quantum entropy).
Sometimes the CMI is denoted  instead by
$S_\rho(\rvx:\rvy|\rve)$,
where $\rho$ is a traced dm-extension of
$\rho_{\rvx,\rvy,\rve}$.
When $N_\rve=1$, CMI degenerates into
the  {\bf mutual information} $S(\rvx:\rvy)$.
Just like classical CMI,
quantum CMI satisfies the chain rule

\beq
S(\rvx:\rvy_1,\rvy_2|\rve)=
S(\rvx:\rvy_1|\rvy_2,\rve)+
S(\rvx:\rvy_2|\rve)
\;.
\label{eq-quant-cmi-chain-rule}
\eeq

Given
$\rho_{\rvx,\rvy}\in dm(\calh_{\rvx,\rvy})$,
the {\bf CMI entanglement} (an information
theoretic measure of quantum entanglement)
is defined as

\beq
E^{CMI}(\rvx:\rvy)=
\frac{1}{2}\inf_{\rho_{\rvx,\rvy,\rve}\in K}(
S_{\rho_{\rvx,\rvy,\rve}}(\rvx:\rvy|\rve))
\;,
\eeq
where the infimum (a generalized minimum) is taken over the
set $K$ of all density matrices
$\rho_{\rvx,\rvy,\rve}\in dm(\calh_{\rvx,\rvy,\rve})$
such that $\tr_{\rve}\rho_{\rvx,\rvy,\rve} = \rho_{\rvx,\rvy}$.
Sometimes, the CMI entanglement is denoted instead
by $E^{CMI}(\rho_{\rvx,\rvy})$, or
by $E^{CMI}_\rho(\rvx:\rvy)$,
where $\rho$ is a traced dm-extension of
$\rho_{\rvx,\rvy}$.
CMI entanglement is also known by the less
scientific name of ``squashed entanglement".
For more information about CMI entanglement,
see Ref.\cite{wiki-cmi-ent}.

If we apply the definition of CMI entanglement
to the right hand side of
Eq.(\ref{eq-quant-cmi-chain-rule}),
we get

\beq
S(\rvx:\rvy_1,\rvy_2|\rve)\geq
2E^{CMI}(\rvx:\rvy_1) +
2E^{CMI}(\rvx:\rvy_2)
\;.
\eeq
Now we are free to
apply the definition of CMI entanglement to
the left hand side of the
previous equation to get:

\beq
E^{CMI}(\rvx:\rvy_1,\rvy_2)\geq
E^{CMI}(\rvx:\rvy_1) +
E^{CMI}(\rvx:\rvy_2)
\;.
\label{eq-synerg}
\eeq
Eq.(\ref{eq-synerg})
can be called {\bf super-additivity of the right
side argument of $E^{CMI}$}. Since entanglement is symmetric
(i.e., $E(\rvx:\rvy)=E(\rvy:\rvx)$),
there is also {\bf
super-additivity of the left side argument $E^{CMI}$}.
Eq.(\ref{eq-synerg}) can also be
called the {\bf synergism of entanglement},
because the whole has more entanglement than
the sum of its parts. If the inequality
in Eq.(\ref{eq-synerg}) were in the opposite
direction, we could call it {\bf sub-additivity} or
{\bf anti-synergism}.

\section{Meta Density Matrix and
\\Purification of a Density Matrix}

In this section, we define meta density matrices,
and purifications of density matrices.
We show that any density matrix
has a purification.

A pure density matrix
$\mu\in dm(\calh_\rvxdot)$
of the form
$\mu = \proj(\sum_{x.} A(x.)\ket{x.})$
will be called a {\bf meta density matrix}.
If $A(x.)$ is the full joint
amplitude associated with a
Bayesian or Markov network $\tilde{G}$,
we will call $\mu$ the {\bf meta density matrix
of the network $\tilde{G}$}.

Suppose $J\subset \zn$ and $J^c=\zn-J$.
Given a density matrix
$\rho\in dm(\calh_{\rvxvec{J}})$,
we will call any pure density matrix
$\mu \in dm(\calh_{\rvxdot})$
such that $\tr_\rvxvec{J^c}(\mu)= \rho$,
a {\bf traced purification of $\rho$}.
More generally, if
$\rho=\Omega(\mu)$ where
the operator $\Omega$ is not
a trace operator,
we
will call $\mu$ a {\bf generalized purification
of $\rho$}.

Crucial to this paper is the
well known fact that any density matrix
has a traced purification. Next, we
will present a proof of this fact. Our
proof is a nice showcase of
Bayesian net ideas and of our notation.

Consider any $\rho\in dm(\calh_\rvx)$.
Let

\beq
\rho = \sum_{x,x'}
\rho(x,x')\ket{x}\bra{x'}
\;.
\eeq
Let $M$ be the matrix
with entries $\rho(x,x')$,
where $x\in St_\rvx$ labels its rows
and $x'\in St_\rvx$ its columns.
$M$ is a Hermitian matrix so it can
be diagonalized. Let
$M = U D U^\dagger$,
where $U$ is a unitary matrix, and $D$
is a real, diagonal matrix.
Set
$U_{x,j} = A(x|j)$ and
$D_{j,j} = |A|^2(j)$,
where $N_\rvj=N_\rvx$.
Then

\beq
\rho(x,x') = \sum_j A(x|j) |A|^2(j) A^*(x'|j)
\;.
\eeq
If we define

\beq
\mu=
\sum_{x,j} A(x|j)A(j)\ket{x,j}
\;,
\eeq
then

\beq
\rho = \tr_\rvj \;\proj(\mu)
\;.
\eeq
A Bayesian net
representation of the previous equation is
\beq
\rho \;\;= \;\;(\rvx)\larrow\stackrel{\tr}{(\rvj)}
\;.
\eeq
The $\tr$ over the $\rvj$
is intended to indicated that
node $\rvj$ should be traced over.
Note that the eigenvectors of $\rho$ become the
transition amplitudes of node $\rvx$,
whereas the square root of the eigenvalues
of $\rho$
become the amplitudes of node $\rvj$.

\section{Measurements of the Meta Density Matrix}

We've shown that any density matrix $\rho$
has a traced purification $\mu$.
Thus, without loss of generality,
we need only consider meta density matrices
$\mu$ and those
density matrices obtained by applying
measurement operators to $\mu$.
In this section, we describe a ``complete"
set of measurement operators that can be
applied to a meta density matrix to obtain
all measurable probabilities codified
within it.

First, consider classical physics.
In particular, consider
$N$ random variables $\rvxdot$
described by a probability distribution $P(x.)$.
Suppose

\beq
\zn=Z_{vis}\cup Z_{sum}
\;,
\label{eq-zn-is-vis-and-sum}
\eeq
 where
$Z_{vis}$ and $Z_{sum}$
are disjoint sets.
Here ``vis" stands for ``visible" and
``sum" for ``summed".
The probability
that
$\rvxvec{Z_{vis}}=\xvec{Z_{vis}}$
is defined as

\beq
P(\xvec{Z_{vis}}) = \sum_{\xvec{Z_{sum}}} P(x.)
\;.
\eeq
$Z_{vis}$  can also be spilt into two
parts. Let

\beq
Z_{vis}=Z_{post}\cup Z_{pre}
\;,
\label{eq-zvis-parts}
\eeq
where
$Z_{post}$ and $Z_{pre}$
are disjoint sets.
The conditional probability
that
$\rvxvec{Z_{post}}=\xvec{Z_{post}}$
given
$\rvxvec{Z_{pre}}=\xvec{Z_{pre}}$
is defined as

\beq
P(\xvec{Z_{post}}|\xvec{Z_{pre}}) =
\frac{
P(\xvec{Z_{post}},\xvec{Z_{pre}})
}{
P(\xvec{Z_{pre}})
}
\;.
\eeq
The conditional expected value
(a.k.a. conditional expectation) of any complex valued
function $f(\cdot)$ of the random
variable $\rvxvec{Z_{vis}}$
is defined as:

\beq
E[f(\rvxvec{Z_{vis}})|
\rvxvec{Z_{pre}}=\xvec{Z_{pre}}]
=
\sum_{\xvec{Z_{post}}}f(\xvec{Z_{vis}})
P(\xvec{Z_{post}}|\xvec{Z_{pre}})
\;.
\label{eq-cl-expect}
\eeq

\begin{figure}[h]
    \begin{center}
    \epsfig{file=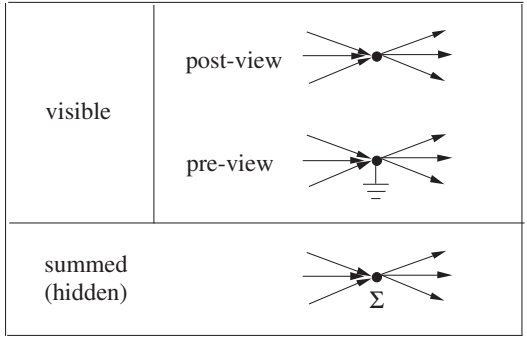, height=1.75in}
    \caption{Various node decorations,
    used with both classical and quantum Bayesian networks,
    to indicate visible and hidden nodes. In a
    probability $P(a|e)=\sum_h P(a,h|e)$,
    $h$ is hidden, $a,e$ are visible, $e$ is pre-viewed
    and $a$ is post-viewed.}
    \label{fig-slot-symbols}
    \end{center}
\end{figure}
Visible (either pre or post viewed)
 and hidden nodes will be indicated on
 a Bayesian network by the node decorations
 show in Fig.\ref{fig-slot-symbols}

Next, consider quantum physics.
In particular,
consider $N$ random variables $\rvxdot$
described by a pure state

\beq
\ket{\phi_{meta}}= \sum_{x.} A(x.) \ket{x.}
\;,
\eeq
or, equivalently, by the meta density matrix

\beq
\mu = \ket{\phi_{meta}}\bra{\phi_{meta}}
=\proj(\ket{\phi_{meta}})
\;.
\eeq
Our next goal is to generalize the
classical physics definitions
Eqs.(\ref{eq-zn-is-vis-and-sum})
to (\ref{eq-cl-expect})
to quantum physics.
Let

\beq
\zn = Z_{vis} \cup Z_{sum}\;,\;\;
Z_{sum}=Z_{Asum} \cup Z_{Psum}
\;,
\eeq
where $Z_{vis}$, $Z_{Asum}$ and
$Z_{Psum}$ are disjoint sets. Here
``Asum" stands
for ``amplitude summed" and
``Psum" stands
for ``probability summed".
The probability
that
$\rvxvec{Z_{vis}}=\xvec{Z_{vis}}$
is defined as

\beq
P(\xvec{Z_{vis}})_{\backslash\rvxvec{Z_{Psum}}} =
\frac{
\sum_{\xvec{Z_{Psum}}}|\sum_{\xvec{Z_{Asum}}} A(\xvec.)|^2
}{
\sum_{\xvec{Z_{vis}}}\; numerator
}
\;.
\label{eq-q-pvis-amp}
\eeq
Note that, contrary to the classical physics case,
this probability depends on
which random variables are summed
coherently (A summed) and
which are summed incoherently
(P summed). We've
indicated this dependence by
the subscript $\backslash\rvxvec{Z_{Psum}}$.
The backslash in this notation is intended
to evoke a mental picture of the diagonal of a matrix,
because the variables that are P summed are
``diagonalized"(
why we say these variables are diagonalized
will become clear to the reader later on,
once he sees Eq.(\ref{eq-qprob-diag-op}) ).
As in the classical physics case, let

\beq
Z_{vis}=Z_{post}\cup Z_{pre}
\;,
\eeq
where
$Z_{post}$ and $Z_{pre}$
are disjoint sets.
The conditional probability
that
$\rvxvec{Z_{post}}=\xvec{Z_{post}}$
given
$\rvxvec{Z_{pre}}=\xvec{Z_{pre}}$
is defined, in analogy to
the classical physics case, by

\beq
P(\xvec{Z_{post}}|\xvec{Z_{pre}})
_{\backslash\rvxvec{Z_{Psum}}} =
\frac{
P(\xvec{Z_{post}},\xvec{Z_{pre}})
_{\backslash\rvxvec{Z_{Psum}}}
}{
P(\xvec{Z_{pre}})
_{\backslash\rvxvec{Z_{Psum}}}
}
\;.
\label{eq-q-cond-prob}
\eeq
Consider a Hermitian operator $\Omega_{\rvxvec{Z_{vis}}}$
acting on $\calh_{\rvxvec{Z_{vis}}}$.
Suppose $\{\ket{\xvec{Z_{vis}}}:
\forall \xvec{Z_{vis}}\}$ are the eigenstates
of $\Omega_{\rvxvec{Z_{vis}}}$, so that

\beq
\Omega_{\rvxvec{Z_{vis}}} =
\sum_{\xvec{Z_{vis}}}\lam_{\xvec{Z_{vis}}}
\ket{\xvec{Z_{vis}}}\bra{\xvec{Z_{vis}}}
\;.
\eeq
In analogy to the classical physics case,
one defines the conditional expected value
of $\Omega_{\rvxvec{Z_{vis}}}$ by

\beq
E[\Omega_{\rvxvec{Z_{vis}}}|
\rvxvec{Z_{pre}}=\xvec{Z_{pre}}]
_{\backslash\rvxvec{Z_{Psum}}}
=
\sum_{\xvec{Z_{post}}}\lam_{\xvec{Z_{vis}}}
P(\xvec{Z_{post}}|\xvec{Z_{pre}})
_{\backslash\rvxvec{Z_{Psum}}}
\;.
\eeq

At this point, we have achieved our
 goal of generalizing the
classical physics definitions
Eqs.(\ref{eq-zn-is-vis-and-sum}) to (\ref{eq-cl-expect})
to quantum physics.
In doing so, we've introduced the
probability
$P(\xvec{Z_{post}}|\xvec{Z_{pre}})
_{\backslash\rvxvec{Z_{Psum}}}$.
The rest of this section will be devoted
to explaining how this probability
can be measured.

To measure
$P(\xvec{Z_{post}}|\xvec{Z_{pre}})
_{\backslash\rvxvec{Z_{Psum}}}$
instead of $P(\xvec{Z_{vis}})
_{\backslash\rvxvec{Z_{Psum}}}$,
one restricts the
range of the random variable
$\rvxvec{Z_{pre}}$ to the single
value $\xvec{Z_{pre}}$.
Of course, one must also divide (``normalize") the
restricted meta density matrix by a constant
so that its trace remains 1.
Next, we show how to measure
$P(\xvec{Z_{vis}})
_{\backslash\rvxvec{Z_{Psum}}}$.

Note that
$P(\xvec{Z_{vis}})_{\backslash\rvxvec{Z_{Psum}}}$
given by Eq.(\ref{eq-q-pvis-amp})
can be expressed as the
expected value, in the meta density matrix $\mu$,
of a projection operator
$\pi^{(a)} \pi^{(b)}\pi^{(c)}$:

\beq
P(\xvec{Z_{vis}})_{\backslash\rvxvec{Z_{Psum}}} =
\frac{
\tr_{\rvxdot}(
\pi^{(a)} \pi^{(b)}\pi^{(c)}\mu)
}{
\sum_{\xvec{Z_{vis}}}\; numerator
}
\;.
\label{eq-quant-prob-as-exp-of-3projs}
\eeq
The projection operator
$\pi^{(a)} \pi^{(b)}\pi^{(c)}$
consists  of
a product of 3 mutually commuting projection operators
defined by

\beq
\pi^{(a)} =
\proj(\ket{\xvec{Z_{vis}}})
\;,
\eeq

\beq
\pi^{(b)} =
\proj(\ket{AV \;\rvxvec{Z_{Asum}}})
\;,
\eeq
and

\beq
\pi^{(c)} =
\sum_{\xvec{Z_{Psum}}}
\proj(\ket{\xvec{Z_{Psum}}})
\;.
\eeq
In $\pi^{(b)}$, we use the ``average"  state vector
$\ket{ AV\;\rvxvec{J}}$, for $J\subset\zn$.
This vector is defined as

\beq
\ket{ AV\; \rvxvec{J}}
= \frac{
\sum_{\xvec{J}} \ket{\xvec{J}}
}{
\sqrt{|J|}
}
\;.
\eeq
The fact that
$P(\xvec{Z_{vis}})_{\backslash\rvxvec{Z_{Psum}}}$
can be expressed as an expected value
of a projection operator suggests one way
of measuring it.

\begin{figure}[h]
    \begin{center}
    \epsfig{file=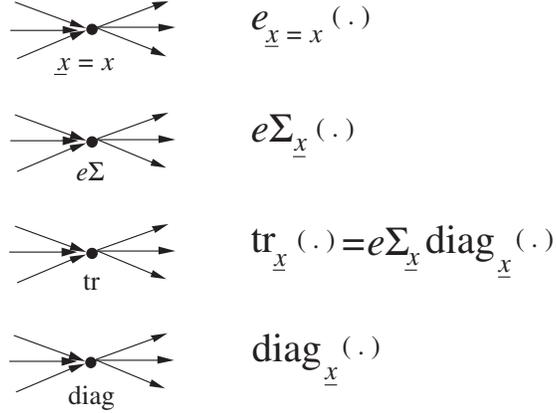, height=2.25in}
    \caption{Various node decorations
    used with quantum Bayesian networks to indicate
    operators acting on the meta density matrix associated
    with the network.}
    \label{fig-meas-symbols}
    \end{center}
\end{figure}

Suppose $\Omega_{\rvx,\rvy}$
is an operator acting on $\calh_{\rvx,\rvy}$.
It is convenient at this point to define
the following super-operators acting
on $\Omega_{\rvx,\rvy}$:

\beq
e_{\rvx=x}(\Omega_{\rvx,\rvy})
 = \bra{x}\Omega_{\rvx,\rvy}\ket{x}
\;,\;\;\;{\rm (entry)}
\eeq

\beq
e\Sigma_\rvx (\Omega_{\rvx,\rvy}) =
\sum_{x,x'}
\bra{x}\Omega_{\rvx,\rvy}\ket{x'}
\;,\;\;\;{\rm (entry\;sum)}
\eeq

\beq
\tr_\rvx (\Omega_{\rvx,\rvy}) =
\sum_{x}
\bra{x}\Omega_{\rvx,\rvy}\ket{x}
\;,\;\;\;{\rm (trace)}
\eeq

\beq
\diag_\rvx (\Omega_{\rvx,\rvy}) =
\sum_{x} \ket{x}\bra{x}\;
\bra{x}\Omega_{\rvx,\rvy}\ket{x}
\;.\;\;\;{\rm (diagonal\;matrix)}
\eeq
We've shown in parenthesis on the right
hand side what we call these operators.
Note that $\diag_\rvx\Omega_{\rvx,\rvy}$
diagonalizes $\Omega_{\rvx,\rvy}$ partially.
$\diag_{\rvx,\rvy} \Omega_{\rvx,\rvy}$
diagonalizes it fully.
\footnote{
Previously, we defined
$\diag(\cdot)$ to be
a function that takes a
vector $\vec{x}$ and returns a diagonal
matrix with $\vec{x}$ along its
diagonal. Here we are defining a different
$\diag(\cdot)$ function. Both
of these
functions return a diagonal matrix,
but they have different domains.
We will use the symbol
$\diag(\cdot)$ for both
of these functions. Which function
we mean will be clear from the context. }
Note that
$\omega\; \diag_\rva = \omega$
for
$\omega = \diag_\rva, \tr_\rva, e_{\rva=a}$.
On the other hand,

\beq
e\Sigma_\rva \diag_\rva = \tr_\rva
\;.
\label{eq-esig-diag}
\eeq
Fig.\ref{fig-meas-symbols} gives node decorations
that will be used to indicate
these operators when acting on
a Bayesian network.

In Eq.(\ref{eq-quant-prob-as-exp-of-3projs}),
we obtained
$P(\xvec{Z_{vis}})_{\backslash\rvxvec{Z_{Psum}}}$
as an expected value of a projection operator.
Alternatively,
$P(\xvec{Z_{vis}})_{\backslash\rvxvec{Z_{Psum}}}$
can be obtained by successive applications
of the operators
$e()$, $e\Sigma()$, $\tr()$, and $\diag()$ to $\mu$:

\bsub
\label{eq-quant-prob-as-meas-of-mu}
\beqa
P(\xvec{Z_{vis}})_{\backslash\rvxvec{Z_{Psum}}} &=&
\frac{
e_{\rvxvec{Z_{vis}}=\xvec{Z_{vis}}}
e\Sigma_{\rvxvec{Z_{Asum}}}
\tr_\rvxvec{Z_{Psum}}
\mu
}{
\sum_{\xvec{Z_{vis}}}\;numerator
}\\
&=&
\frac{
e_{\rvxvec{Z_{vis}}=\xvec{Z_{vis}}}
e\Sigma_{\rvxvec{Z_{Asum}},\rvxvec{Z_{Psum}}}
\diag_\rvxvec{Z_{Psum}}
\mu
}{
\sum_{\xvec{Z_{vis}}}\;numerator
}
\;.
\label{eq-qprob-diag-op}
\eeqa
\esub
Eq.(\ref{eq-qprob-diag-op})
follows from Eq.(\ref{eq-esig-diag}).
Here, the operators
$e()$, $e\Sigma()$, $\tr()$, and $\diag()$
can be interpreted as
measurements (or lack thereof)
of the density matrix
they act upon.\footnote{The software program Quantum Fog can
calculate
$P(\xvec{Z_{post}}|\xvec{Z_{pre}})
_{\backslash\rvxvec{Z_{Psum}}}$
numerically.
Conditioning on
$\rvxvec{Z_{pre}}=\xvec{Z_{pre}}$ is
already implemented in the current version, 2.0,
of Quantum Fog; it
corresponds to allowing only one ``active"
state for
each of the nodes $\rvx_j$ for $j\in Z_{pre}$.
On the other hand,
only a special case of
the  distinction
between P-summed and A-summed
is implemented in version 2.0. In version 2.0,
$\rvxvec{Z_{Psum}}$ is always assumed to equal
the set of external nodes
minus the set of visible ones. More general
sets $\rvxvec{Z_{Psum}}$
will be implemented in future
versions of Quantum Fog.}

In Eqs.(\ref{eq-quant-prob-as-meas-of-mu}),
$\tr_\rvxvec{Z_{Psum}}$ means
observe (=measure) the random
variable $\rvxvec{Z_{Psum}}$,
and then forget the outcome.
$e_{\rvxvec{Z_{vis}}=\xvec{Z_{vis}}}$
means measure of the random
variable $\rvxvec{Z_{vis}}$ once.
$e\Sigma_{\rvxvec{Z_{sum}}}$
means do no
observe the random variable
$\rvxvec{Z_{sum}}$.
It remains for us to interpret
$\diag_{\rvxvec{Z_{Psum}}}$ as a measurement.

For any density matrix
$\rho_{\rvx\rvy}\in dm(\calh_{\rvx,\rvy})$,
the operator $\diag_\rvx$
is what is called a von Neumann measurement.
It can be implemented physically in two steps:
(1) measure the random variable $\rvx$;
if the outcome is $x$, emit $\ket{x}\bra{x}$,
and (2)
repeat the measurement
many times, without
discriminating on any of the outcomes
(mathematically, this corresponds to
summing over the
outcomes $x$ of the measurements).

A second way of implementing
$\diag_\rvx$ is as follows.
The Bayesian net
\beq(\rvx)\larrow(\rvy)\eeq
with transition matrix $A(x|y)A(y)$
can be replaced by a
Bayesian net
\beq(\rvx')\larrow(\rvx)\larrow(\rvy)\eeq
with transition matrix
$A(x'|x)A(x|y)A(y)$,
where
$A(x|x')=e^{i\theta_x}\delta(x',x)$
$\forall x,x'\in St_\rvx$.
Assume that the variables $\{\theta_x: \forall x\}$
are i.i.d. (independent, identically distributed)
classical random variables, and each is uniformly
distributed over $[0, 2\pi]$. Let an overline
denote an average over these variables.
An effect of adding the node $\rvx'$
to the network
is that we must replace

\beq
\rho_{\rvx,\rvy}=\sum_{x,y,x',y'}
\rho_{xy,x'y'}\ket{xy}\bra{x'y'}
\;
\eeq
by

\beq
\rho_{\rvx^\theta,\rvy}=\sum_{x,y,x',y'}
\rho_{xy,x'y'}
e^{i\theta_x}\ket{xy}
\bra{x'y'}e^{-i\theta_{x'}}
\;.
\eeq
Clearly,

\beq
\overline{
\rho_{\rvx^\theta, \rvy}
}=
\diag_\rvx \rho_{\rvx,\rvy}
\;,
\eeq
and
\footnote{Of course, for an arbitrary
polynomial function $f$, one has
$
\overline{
f(
\rho_{\rvx^\theta, \rvy}
)
}
\neq
f(
\overline{
\rho_{\rvx^\theta, \rvy}
}
)
$,
but this is not a show stopper, since the density
matrix only enters linearly in
the formula for the expected value
of any observable.}

\beq
\overline{
\tr_\rvx[\Omega\;
\rho_{\rvx^\theta, \rvy}]
}
=
\tr_\rvx[\Omega\;
\overline{
\rho_{\rvx^\theta, \rvy}}]
=
\tr_\rvx[\Omega\;
\diag_\rvx \rho_{\rvx,\rvy}]
\;,
\eeq
for any operator $\Omega$ acting
on $\calh_{\rvx,\rvy}$.
Thus, the operator
$\diag_\rvx$ can be implemented
physically merely by taking many measurements
for which $\theta_x$ varies
randomly.

A third way of implementing $\diag_\rvx$
is by adding an additional
node that is traced over.
For example, suppose
$\rho_\rvx\in dm(\calh_\rvx)$
can be expressed in the form

\beq
\rho_\rvx = \diag_\rvx (\mu)
\;,\;\;
\mu = \proj(\sum_x A(x)\ket{x})
\;.
\label{eq-diag-begone-bef-x}
\eeq
We can introduce a node $\rvj$ such that
$St_\rvj=St_\rvx$ and $A(j) = A(\rvx= j)$.
Then

\beq
\rho_\rvx = tr_{\rvj}(\tilde{\mu})
\;,\;\;
\tilde{\mu}=
\proj(
\sum_{x,j}
\delta^x_{j} A(j)\ket{x,j})
\;.
\label{eq-diag-begone-aft-x}
\eeq
$\mu$ is a generalized purification of $\rho_\rvx$
whereas
$\tilde{\mu}$ is a traced one.
By expressing $\rho_\rvx$ in
terms of $\tilde{\mu}$ instead of $\mu$,
we get rid of the $\diag_\rvx$
operator at the expense of adding
an additional node $\rvj$
that
we trace over. A Bayesian network
representation of
the essence of
Eqs.(\ref{eq-diag-begone-bef-x})
and
(\ref{eq-diag-begone-aft-x})
is:

\beq
\stackrel{\diag}{(\rvx)}\;\;=\;\;
(\rvx)\larrow \stackrel{\tr}{(\rvj)}
\;.
\eeq
As a more general example of this
method of implementing $\diag_{\rvx}$,
suppose
$\rho_{\rvy,\rvx}\in dm(\calh_{\rvy,\rvx})$
can be expressed in the form

\beq
\rho_{\rvy,\rvx} =
\diag_\rvx (\mu)
\;,\;\;
\mu = \proj(\sum_{y,x}A(y|x)A(x)\ket{y,x})
\;.
\label{eq-diag-begone-bef-yx}
\eeq
Once again,  introduce a node
 $\rvj$ such that
$St_\rvj=St_\rvx$ and $A(j) = A(\rvx= j)$.
Then

\beq
\rho_{\rvy,\rvx} =
\tr_{\rvj} (\tilde{\mu})
\;,\;\;
\tilde{\mu} = \proj(\sum_{y,x, j}
A(y|x)\delta^x_{j} A(j)\ket{y,x, j})
\;,
\label{eq-diag-begone-aft-yx}
\eeq
A Bayesian network
representation of
the essence of
Eqs.(\ref{eq-diag-begone-bef-yx})
and
(\ref{eq-diag-begone-aft-yx})
is:

\beq
(\rvy)\larrow\stackrel{\diag}{(\rvx)}\;\;=\;\;
(\rvy)\larrow(\rvx)\larrow \stackrel{\tr}{(\rvj)}
\;.
\eeq

The Schmidt Decomposition
is very popular in the
Quantum Information Theory literature.
As an illustration of the
use of the entry-sum operator $e\Sigma$,
let us consider
the Schmidt Decomposition from the
point of view of Bayesian networks.
The Schmidt Decomposition
is the statement that given a pure state
$\mu_1\in dm(\calh_{\rvx,\rvy})$ of the form

\beq
\mu_1 = \proj(
\sum_{x,y} A(x,y) \ket{x,y})
\;,
\eeq
the coefficients $A(x,y)$
can be expressed in the form

\beq
A(x,y) =
\sum_j  A(x|j) A(y|j)A(j)
\;,
\label{eq-basic-schmidt}
\eeq
where $A(j)\geq 0\;\forall j$,
$\sum_j |A(j)|^2=1$,
$\sum_x |A(x|j)|^2=1\;\forall j$,
$\sum_y |A(y|j)|^2=1\;\forall j$.

The fact that any
$A(x,y)$ can be expressed in the
 form given by Eq.(\ref{eq-basic-schmidt})
is a re-statement of the Singular Value
Decomposition Theorem. This is why.
Let $M$ be the matrix
with entries $A(x,y)$,
where $x\in St_\rvx$ labels its rows
and $y\in St_\rvy$ its columns.
According to the Singular Value Decomposition
theorem, $M$ can be expressed in the form
$M = U D V^\dagger$,
where $U$ and $V$ are unitary matrices
and $D$ is a non-negative, diagonal matrix.
If we let
$U_{x,j} = A(x|j)$,
$D_{j,j} = A(j)$,
$V^*_{y,j} = A(y|j)$,
then Eq.(\ref{eq-basic-schmidt}) follows.

To obtain a Bayesian net
picture of the Schmidt Decomposition, note that
if we define $\mu_2\in dm(\calh_{\rvx,\rvy,\rvj})$
by

\beq
\mu_2 = \proj(
\sum_{x,y,j} A(x|j) A(y|j)A(j)\ket{x,y,j})
\;,
\eeq
then

\bsub
\beqa
e\Sigma_{\rvj} (\mu_2)&=&
\proj(\sum_{x,y,j} A(x|j) A(y|j)A(j)\ket{x,y})\\
&=& \mu_1
\;.
\label{eq-bnet-schmidt}
\eeqa
\esub

\begin{figure}[h]
    \begin{center}
    \epsfig{file=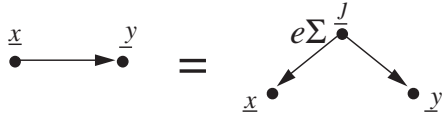, height=.7in}
    \caption{Bayesian net representation of
    the Schmidt Decomposition, as given by
    Eq.(\ref{eq-bnet-schmidt}).}
    \label{fig-schmidt}
    \end{center}
\end{figure}
Eq.(\ref{eq-bnet-schmidt}) is illustrated
by Fig.\ref{fig-schmidt}.

Eq.(\ref{eq-bnet-schmidt})
gives an example of the use of the
entry-sum operator $e\Sigma$. Note
that this operator takes a pure state
of tensor rank $n\geq 2$ into
a pure state of tensor rank $n-1$. Indeed,

\beq
e\Sigma_\rva\; \proj(\sum_{x,a}A(x,a)\ket{x,a})=
\proj( \sum_{x,a}A(x,a)\ket{x})
\;.
\eeq
$e\Sigma$ also takes a pure state of tensor rank
$n=1$ into a non-negative number. Indeed,
for $\ket{\psi}\in\calh_\rva$,

\beq
e\Sigma_\rva \ket{\psi}\bra{\psi} =
\left|\sum_{a}\av{a|\psi}\right|^2
\;.
\eeq

Note that when $N_\rva=1$,
the entry-sum operator $e\Sigma_\rva$
equals
the entry operator
$e_{\rva=a}$. Thus,
$e_{\rva=a}$
can be viewed as a special case of
$e\Sigma_\rva$.
It's clear that $e_{\rva=a}$ inherits
from $e\Sigma_\rva$ the
property that:
it takes a pure state
of tensor rank $n\geq 2$ into
a pure state of tensor rank $n-1$,
and it takes a pure state of tensor rank
$n=1$ into a non-negative number.

Suppose $\mu \in dm(\calh_{\rvxvec{Z_{1,N}}})$ is a
pure density matrix,
and $\rho$ is a density matrix, and
$\rho = (\prod_{j\in J}\omega_{\rvx_j})\mu$,
where $J\subset Z_{1,N}$ and
$\omega_{\rvx_j}
\in
\{
e_{\rvx_j=x_j},
e\Sigma_{\rvx_j},
\tr_{\rvx_j},
\diag_{\rvx_j}\}$.
We've shown  that
$e_{\rvx_j=x_j}$,
and $e\Sigma_{\rvx_j}$ both
take a pure density matrix to
another pure density matrix, so
one can easily find
a pure density matrix
$\mu' \in dm(\calh_{\rvxvec{Z_{1,N'}}})$
such that
$\rho = (\prod_{j\in J'}\omega_{\rvx_j})\mu'$,
where $J'\subset Z_{1,N'}$ and
$\omega_{\rvx_j}
\in
\{
\tr_{\rvx_j},
\diag_{\rvx_j}\}$.
We've shown that
each operator $\diag_{\rvx_j}$
can be traded for an extra node
that is traced over.
Thus,
one can easily find
a pure density matrix
$\mu'' \in dm(\calh_{\rvxvec{Z_{1,N''}}})$
such that
$\rho = (\prod_{j\in J''}\tr_{\rvx_j})\mu''$,
where $J''\subset Z_{1,N''}$.
To summarize, given a
generalized (i.e, made with entry, entry-sum,
trace and diag operators) purification
of $\rho$, one
can easily find a traced purification
of $\rho$. A generalized purification of $\rho$ might be
convenient for certain purposes, but
not for others. Luckily, it can be easily
replaced by a traced one.

\section{Conditional Amplitudes}
\label{sec-cond-amp}

In this section, we define
conditional amplitudes. These are
a natural generalization of conditional
probabilities.

Consider a meta density matrix
$\mu = \proj(\sum_{x.} A(x.)\ket{x.})$
Its complex amplitude $A(x.)$
can be parameterized as

\beq
A(x.) = e^{i\theta(x.)}\rootp(x.)
\;,
\eeq
where the $\theta(x.)$ are real
and $P\in pd(St_\rvxdot)$.
Choose an
arbitrary state  of $(\rvxdot)$,
and call it the {\bf reference state}
$(x.^o)$\;\;.
It is convenient to constrain
$\theta(x.)$
by assuming that it vanishes
at the reference state:

\beq
\theta(x.^o)=0
\;\;\;.
\eeq
For $J\subset \zn$ and $J^c= \zn-J$,
we define

\begin{subequations}
\beq
\theta(\xvec{J}) = \theta(\xvec{J}, \xovec{J^c})
\;,
\eeq

\beq
P(\xvec{J})= \sum_{\xvec{J^c}}P(x.)
\;,
\eeq
and

\beq
A(\xvec{J}) =
e^{i\theta(\xvec{J})}
\rootp(\xvec{J})
\;.
\eeq
\end{subequations}
For disjoint sets $J_1, J_2\subset \zn$,
we define

\begin{subequations}
\beq
\theta(\xvec{J_1}|\xvec{J_2})=
\theta(\xvec{J_1},\xvec{J_2})-
\theta(\xvec{J_2})
\;,
\eeq

\beq
P(\xvec{J_1}|\xvec{J_2})=
\frac{P(\xvec{J_1},\xvec{J_2})}
{P(\xvec{J_2})}
\;,
\eeq
and

\beq
A(\xvec{J_1}|\xvec{J_2})=
\frac{A(\xvec{J_1},\xvec{J_2})}
{A(\xvec{J_2})}
\;.
\eeq
\end{subequations}
Note that
\beq
\theta(\xovec{J_1}|\xvec{J_2})=0
\;,
\eeq

\beq
phase( \bra{x.}\mu\ket{y.}) =
\theta(x.)-\theta(y.)
\;,
\eeq
and

\beq
\theta(x.) = phase( \bra{x.}\mu\ket{x.^o})
\;.
\eeq

\section{Probabilistic Conditional Independence}

This section, divided into 3 subsections,
explores the notion of conditional independence
in both classical and quantum physics.

Henceforth, by an
{\bf
independency},
we will mean a triplet
$(\rvxvec{J} \perp \rvxvec{K}| \rvxvec{E})$,
where $J,K,E\subset \zn$ are disjoint.
(If $J$ and $K$ are disjoint but
overlap with $E$, replace
$(\rvxvec{J} \perp \rvxvec{K}| \rvxvec{E})$
by $(\rvxvec{J-E} \perp \rvxvec{K-E}| \rvxvec{E})$).
If the sets $J$ and $K$
both contain more than one element,
we will call it a {\bf global independency}.
If $|J|+|K|+|E|=N$, we will say
that $(\rvxvec{J} \perp \rvxvec{K}| \rvxvec{E})$
is an {\bf all-encompassing independency}.
We will use the
word {\bf I-set} as an abbreviation
for ``independencies set"; that is, a set
whose members are independencies.
It is convenient to introduce a symbol
for the set of all possible independencies:

\beq
\cali(\rvxdot)=
\{I:I = (\rvxvec{J} \perp \rvxvec{K}| \rvxvec{E});
J,K,E\subset \zn\;{\rm are\;disjoint}\;\}
\;.
\eeq

\subsection{Types of Probabilistic Conditional Independence}

In this section, we define
classical
conditional independence and three
quantum analogues of it,
type-A,
type-CMI,
and type-\cmip.

Consider first classical physics and probability.
Let $J,K,E\subset \zn$ be disjoint sets. We say
$\rvxvec{J}$ and $\rvxvec{K}$
are {\bf conditionally independent}
 given $\rvxvec{E}$
iff

\beq
P(\xvec{J}, \xvec{K} | \xvec{E})=
P(\xvec{J}| \xvec{E})
P(\xvec{K}| \xvec{E})
\;\;
\forall \xvec{J}, \xvec{K}, \xvec{E}
\;.
\label{eq-prob-cond-indep}
\eeq
Eq.(\ref{eq-prob-cond-indep})
is clearly equivalent to requiring
that

\beq
P(\xvec{J}| \xvec{K},\xvec{E})=
P(\xvec{J}| \xvec{E})
\;,
\eeq
or

\beq
P(\xvec{K}| \xvec{J},\xvec{E})=
P(\xvec{K}| \xvec{E})
\;.\eeq
We define the function
$\tau^P:\cali(\rvxdot)\rarrow Bool$
by the statement:
$\tau^P(\rvxvec{J} \perp \rvxvec{K} | \rvxvec{E})$
is true iff Eq.(\ref{eq-prob-cond-indep}) is true.
Think of $\tau^P$ as a ``truth function" that
decides whether its argument is false=0 or
true=1.

In classical physics,
conditional independence and vanishing CMI
are equivalent. Indeed,

\begin{theo}\label{th-class-cmi}
\beq
H(\rvx:\rvy)=0
\;\;{\rm iff}\;\;
P(x,y)= P(x)P(y)
\;\;\forall x,y
\;,
\eeq
and
\beq
H(\rvx:\rvy|\rve)=0
\;\;{\rm iff}\;\;
P(x,y|e)= P(x|e)P(y|e)
\;\;\forall x,y,e
\;.
\eeq
\end{theo}
\proof
The proof can be found in Ref.\cite{Cover}.
\qed

Now consider quantum physics. Our
goal is to find the quantum counterpart of
Eq.(\ref{eq-prob-cond-indep})
and Theorem \ref{th-class-cmi}. Consider a
meta density matrix
$\mu = \proj(\sum_{x.} A(x.)\ket{x.})$.
Let $J,K,E\subset \zn$ be disjoint sets. We say
$\rvxvec{J}$ and $\rvxvec{K}$
are {\bf type-A conditionally independent}
 given $\rvxvec{E}$ iff

\beq
A(\xvec{J}, \xvec{K} | \xvec{E})=
A(\xvec{J}| \xvec{E})
A(\xvec{K}| \xvec{E})
\;\;\forall \xvec{J}, \xvec{K}, \xvec{E}
\;.
\label{eq-amp-cond-indep}
\eeq
We say
$\rvxvec{J}$ and $\rvxvec{K}$
are {\bf type-CMI conditionally independent}
 given $\rvxvec{E}$
iff
\beq
S_\mu(\rvxvec{J} : \rvxvec{K} | \rvxvec{E})=0
\;.
\label{eq-cmi-cond-indep}
\eeq
(Note that we trace over
all random variables
$\rvx_n$ such that $n\not \in J\cup K\cup E$).
We say
$\rvxvec{J}$ and $\rvxvec{K}$
are {\bf type-\cmip conditionally independent}
 given $\rvxvec{E}$
iff
\beq
S_{\diag_\rvxvec{E}(\mu)}(\rvxvec{J} : \rvxvec{K} | \rvxvec{E})=0
\;.
\label{eq-cmip-cond-indep}
\eeq

We define the function
$\tau^A:\cali(\rvxdot)\rarrow Bool$
by the statement:
$\tau^A(\rvxvec{J} \perp \rvxvec{K} | \rvxvec{E})$
is true iff Eq.(\ref{eq-amp-cond-indep}) is true.
Likewise,
$\tau^{CMI}(\rvxvec{J} \perp \rvxvec{K} | \rvxvec{E})$
iff Eq.(\ref{eq-cmi-cond-indep}).
Likewise,
$\tau^{CMI'}(\rvxvec{J} \perp \rvxvec{K} | \rvxvec{E})$
iff Eq.(\ref{eq-cmip-cond-indep}).

In classical physics, type-A and type-CMI
conditional independence
are equivalent, but
in quantum physics,
neither one implies the other.
We will give counterexamples of
this later. But first,
we will give easy-to-check necessary and
sufficient conditions for
a vanishing quantum CMI.

\begin{theo}
For $\rho_{\rvx\rvy}\in dm(\calh_{\rvx,\rvy})$,

\beq
S_{\rho_{\rvx\rvy}}(\rvx:\rvy)=0
\;\;{\rm iff}\;\;
\rho_{\rvx\rvy}=
\rho_\rvx\rho_\rvy
\;,
\label{eq-q-indep-nec-suf}
\eeq
where
$\rho_\rvx$ and
$\rho_\rvy$
are partial traces of $\rho_{\rvx\rvy}$.
For $\rho_{\rvx\rvy\rve}\in
dm(\calh_{\rvx,\rvy,\rve})$,

\beq
S_{\rho_{\rvx\rvy\rve}}(\rvx:\rvy|\rve)=0
\;\;{\rm iff}\;\;
\rho_{\rvx\rvy\rve}=
\sum_e  \ket{e}\bra{e} w(e)
\rho_\rvx^{(e)}\rho_\rvy^{(e)}
\;,
\label{eq-q-cond-indep-nec-suf}
\eeq
where
$w(\cdot) \in pd(St_\rve)$, and, for all $e$,
$\rho_\rvx^{(e)}\in dm(\calh_\rvx)$,
$\rho_\rvy^{(e)}\in dm(\calh_\rvy)$.
\end{theo}
\proof
Eq.(\ref{eq-q-cond-indep-nec-suf})
implies Eq.(\ref{eq-q-indep-nec-suf}).
Proving $\Larrow$ for
Eq.(\ref{eq-q-cond-indep-nec-suf})
is a simple calculation. It was
pointed out in Ref.
\cite{Tucci-separability}.
Proving $\Rarrow$ for
Eq.(\ref{eq-q-cond-indep-nec-suf})
is much more technical. A weak version
of it was proven in Ref.\cite{Tucci-separability}.
The strong version presented here was
first proven in Ref.\cite{Petz}.
\qed

\begin{theo}\label{th-icmi-iff}
Consider a meta density matrix
$\mu = \proj(\sum_{x.} A(x.)\ket{x.})$.
Suppose $K_1,K_2,E\subset \zn$ are disjoint sets,
 $U=K_1\cup K_2\cup E$,
and $U^c = \zn-U$.
Let
$I=(\rvxvec{K_1} \perp \rvxvec{K_2} | \rvxvec{E})$.
\newline
$\tau^{CMI'}(I)$ iff
$\forall (\xvec{E},\xvec{K_1},\xvec{K_2},\xvec{K_1'},\xvec{K_2'})$

\beq
\sum_{\xvec{U^c}}
\biprod{
A(\xvec{K_1},\xvec{K_2},\xvec{U^c},\xvec{E})
}{
A^*(\xpvec{K_1},\xpvec{K_2},\xvec{U^c},\xvec{E})
}
=
w(\xvec{E})
\rho^{(\xvec{E})}_1(\xvec{K_1},\xpvec{K_1})
\rho^{(\xvec{E})}_2(\xvec{K_2},\xpvec{K_2})
\;,
\label{eq-tau-cmi-p-equiv}
\eeq
where
$w(\cdot)\in pd(St_\rvxvec{E})$,
and where ,
for $j\in\{1,2\}$, $\forall \xvec{E}$,
$\rho^{(\xvec{E})}_j\in dm(\calh_{\rvxvec{K_j}})$.
\end{theo}
\proof

Define $\rho$ by

\bsub
\beqa
\rho &=& \diag_\rvxvec{E} \tr_\rvxvec{U^c} \mu\\
&=& \sum_\xvec{E}
\ket{\xvec{E}}
\bra{\xvec{E}}
\sum_{
\xvec{K_1}\xvec{K_2}
\atop
\xpvec{K_1}\xpvec{K_2}
}
\sum_\xvec{U^c}\biprod{
A(\xvec{K_1}, \xvec{K_2}, \xvec{U^c}, \xvec{E})
}{
A^*(\xpvec{K_1}, \xpvec{K_2}, \xvec{U^c}, \xvec{E})
}
\ket{\xvec{K_1}\xvec{K_2}}
\bra{\xpvec{K_1}\xpvec{K_2}}
\;.\nonumber\\&&
\label{eq-rho-def-for-tau-cmi-p}
\eeqa
\esub
$\tau^{CMI'}(I)$ is equivalent to
$S_\rho(\rvxvec{K_1}:\rvxvec{K_2}|\rvxvec{E})=0$.

Recall that for any $\rho_{sol}\in dm(\calh_\rvxvec{U})$,

\beq
S_{\rho_{sol}}(\rvxvec{K_1}:\rvxvec{K_2}|\rvxvec{E})=0
\;\;{\rm iff}\;\;
\rho_{sol}=
\sum_\xvec{E}
\ket{\xvec{E}}
\bra{\xvec{E}}
w(\xvec{E})
\rho_{\rvxvec{K_1}}^{(\xvec{E})}
\rho_{\rvxvec{K_2}}^{(\xvec{E})}
\;.
\label{eq-s-rho-sol}
\eeq

($\Rarrow$)
By setting $\rho$ equal to $\rho_{sol}$, we
prove Eq.(\ref{eq-tau-cmi-p-equiv}).

($\Larrow$)By plugging Eq.(\ref{eq-tau-cmi-p-equiv})
into Eq.(\ref{eq-rho-def-for-tau-cmi-p}), we
 show that
$\rho$ satisfies the right hand side of
Eq.(\ref{eq-s-rho-sol}), so it satisfies
the left hand side of the same equation.
\qed

We are finally ready to prove that for
type-A and
type-CMI conditional independence,
neither one of these
implies the other.

\begin{theo}
Suppose $K_1,K_2,E\subset \zn$ are disjoint sets, and
$I=(\rvxvec{K_1} \perp \rvxvec{K_2} | \rvxvec{E})$.
$\tau^{CMI}(I)\nRarrow \tau^A(I)$
and
$\tau^{CMI}(I)\nLarrow \tau^A(I)$.
Also,
$\tau^{CMI'}(I)\nRarrow \tau^A(I)$
and
$\tau^{CMI'}(I)\nLarrow \tau^A(I)$.
Also, $\tau^{CMI}(I)\Rarrow \tau^{CMI'}(I)$.
\end{theo}
\proof
Let $U=K_1\cup K_2\cup E$,
and $U^c = \zn-U$.
For our counterexamples, we will assume
$\xvec{K_1}\rarrow x_1$,
$\xvec{K_2}\rarrow x_2$,
$\xvec{U^c}\rarrow a$,
where $x_1,x_2,a$ are Boolean variables.
We will take $N_\rvxvec{E}=1$,
and indications of any dependence on
$\xvec{E}$ will be suppressed.
We will take $(0,0,0)$
to be our reference
state (i.e., the state $(x^o_1,x^o_2,a^o)$
for which $\theta(x^o_1,x^o_2,a^o)=0)$.
We will abbreviate $\theta(x_1,x_2,a)$ by
$\theta_{x_1,x_2,a}$.

$\tau^{CMI}(I)\Rarrow \tau^{CMI'}(I)$ is obvious.
Since we will assume $N_\rvxvec{E}=1$,
our example of
$\tau^{CMI'}(I)\nRarrow \tau^A(I)$
will also prove
$\tau^{CMI}(I)\nRarrow \tau^A(I)$.
Likewise,
our example of
$\tau^{CMI'}(I)\nLarrow \tau^A(I)$
will also prove
$\tau^{CMI}(I)\nLarrow \tau^A(I)$.

({\bf proof of $\tau^{CMI'}\nRarrow\tau^A$}) Assume

\beq
\left\{
\begin{array}{l}
A(x_1,x_2,a) = \delta_{x_1,x_2}^{1,1}
\frac{e^{i\theta_{x_1,x_2,a}}}{\sqrt{2}}
\\
\theta_{x_1,x_2,a}
=
\xi\delta_{x_1,x_2,a}^{1,1,0}
\;,\;\;{\rm where}\;\xi\in \RR-2\pi\ZZ
\end{array}
\right.
\;.
\eeq
This $A(x_1,x_2,a)$
satisfies

\beq
\sum_a
\biprod{
A(x_1,x_2,a)
}{
A^*(x'_1,x'_2, a)
}=
\delta_{x_1,x'_1}^{1,1}
\delta_{x_2,x'_2}^{1,1}
\;.
\eeq
Therefore, $\tau^{CMI'}(I)$ is true.
This  $A(x_1,x_2,a)$
also satisfies

\beq
\left\{
\begin{array}{l}
A(x_1,x_2) =
\delta_{x_1,x_2}^{1,1}
e^{i\theta_{110}}
=\delta^{1,1}_{x_1,x_2}e^{i\xi}
\\
A(x_1) =
\delta_{x_1}^1
e^{i\theta_{100}}
=\delta^1_{x_1}
\\
A(x_2) =
\delta_{x_2}^1
e^{i\theta_{010}}
=\delta^1_{x_2}
\end{array}
\right.
\;.
\eeq
Hence
\beq
A(x_1,x_2)\neq A(x_1)A(x_2)
\;,
\eeq
which means $\tau^A(I)$ is false.

({\bf proof of $\tau^{CMI'}\nLarrow\tau^A$}) Assume

\beq
\left\{
\begin{array}{l}
A(x_1,x_2,a) = \frac{
e^{i\theta_{x_1,x_2,a}}
}{
\sqrt{8}
}
\\
\theta_{x_1,x_2,a}
=
\xi\delta_{x_1,x_2,a}^{1,1,1}
\;,\;\;{\rm where}\;\xi\in \RR-2\pi\ZZ
\end{array}
\right.
\;.
\eeq
This $A(x_1,x_2,a)$
satisfies

\beq
\left\{
\begin{array}{l}
A(x_1,x_2) =
\frac{e^{i\theta_{x_1x_2 0}}}{\sqrt{4}}=
\frac{1}{\sqrt{4}}
\\
A(x_1) =
\frac{e^{i\theta_{x_1 00}}}{\sqrt{2}}=
\frac{1}{\sqrt{2}}
\\
A(x_2) =
\frac{e^{i\theta_{0x_2 0}}}{\sqrt{2}}=
\frac{1}{\sqrt{2}}
\end{array}
\right.
\;.
\eeq
Therefore,

\beq
A(x_1,x_2)= A(x_1)A(x_2)
\;,
\eeq
which means $\tau^A(I)$ is true.
This $A(x_1,x_2,a)$ also
satisfies

\beq
\sum_a
\biprod{
A(x_1,x_2,a)
}{
A^*(x'_1,x'_2, a)
}=
\frac{1 + e^{i\xi[\delta_{x_1,x_2}^{1,1}
-\delta_{x'_1,x'_2}^{1,1}]}}{8}
\;.
\label{eq-a-biprod-eg}
\eeq
Let's show that assuming $\tau^{CMI'}(I)$
leads to a contradiction.
Theorem \ref{th-icmi-iff} implies (i)
and Eq.(\ref{eq-a-biprod-eg})
implies (ii) in the following:

\beq
phase[\rho_1(0,0)\rho_2(1,0)]
\stackrel{(i)}{=}
phase(
\sum_a
\biprod{
A(0,1,a)
}{
A^*(0,0, a)
}
)
\stackrel{(ii)}{=
}0
\;,
\eeq
and

\beq
phase[\rho_1(1,1)\rho_2(1,0)]
\stackrel{(i)}{=}
phase(
\sum_a
\biprod{
A(1,1,a)
}{
A^*(1,0, a)
}
)
\stackrel{(ii)}{=}
phase( 1 + e^{i\xi})
\;.
\eeq
Since $\rho_1(0,0)$ and
$\rho_1(1,1)$ are supposed to
be real, the right hand sides of the
 two previous equations are supposed to
be equal. They aren't---a contradiction.
\qed

There is, however, one subset of $\cali(\rvxdot)$
over
which $\tau^A$ and $\tau^{CMI'}$ agree.

\begin{theo}\label{th-tau-a-tau-cmip-connection}
Suppose $K_1,K_2,E\subset \zn$ are disjoint sets, and
$I=(\rvxvec{K_1} \perp \rvxvec{K_2} | \rvxvec{E})$.
If $|K_1|+|K_2|+|E| = N$, then
$\tau^A(I)= \tau^{CMI'}(I)$.
\end{theo}
\proof
According to Theorem \ref{th-icmi-iff},
$\tau^{CMI'}(I)$ is equivalent to:

\beq
\biprod{
A(\xvec{K_1},\xvec{K_2},\xvec{E})
}{
A^*(\xpvec{K_1},\xpvec{K_2}\xvec{E})
}
=
w(\xvec{E})
\rho^{(\xvec{E})}_1(\xvec{K_1},\xpvec{K_1})
\rho^{(\xvec{E})}_2(\xvec{K_2},\xpvec{K_2})
\;,
\eeq
where $w(\cdot)$ is a probability distribution,
and for $j=1,2$, $\forall \xvec{E}$,
$\rho^{(\xvec{E})}_j$
are density matrices. $\tau^A(I)$,
on the other hand, is equivalent to

\beq
A(\xvec{K_1},\xvec{K_2},\xvec{E})
=
A(\xvec{K_1}|\xvec{E})
A(\xvec{K_2}|\xvec{E})
A(\xvec{E})
\;.
\eeq
Clearly, $\tau^A(I)$ implies $\tau^{CMI'}(I)$.
To show that $\tau^{CMI'}(I)$ implies $\tau^A(I)$,
define $\theta(x.)=phase(A(x.))$,
$|A|(\xvec{E})=\sqrt{w(\xvec{E})}$,
and, for $j=1,2$,
$|A|(\xvec{K_j}|\xvec{E})=
\sqrt{\rho^{(\xvec{E})}_j(\xvec{K_j},\xvec{K_j})}$.
\qed

\subsection{Reduction and Combination Rules
for Independencies}

Consider the following {\bf reduction and
combination
rules} for independencies:
\begin{enumerate}
\renewcommand{\labelenumi}{(\alph{enumi})}

\item {\bf (Decomposition/$2\rarrow 1$)}\newline
$\tau^\eta(\rvx \perp \rvy_1,\rvy_2 | \rve)
\Rarrow
\tau^\eta(\rvx \perp \rvy_2 | \rve)$

\item {\bf (Weak Union/$2\rarrow 1'$)}\newline
$\tau^\eta(\rvx \perp \rvy_1,\rvy_2 | \rve)
\Rarrow
\tau^\eta(\rvx \perp \rvy_1 | \rvy_2,\rve)$

\item {\bf (Contraction/$1',1\rarrow 2$)}\newline
$\tau^\eta(\rvx \perp \rvy_1 | \rvy_2,\rve)$
and
$\tau^\eta(\rvx \perp \rvy_2 | \rve)
\Rarrow
\tau^\eta(\rvx \perp \rvy_1,\rvy_2 | \rve)$

\item {\bf (Intersection/$1',1'\rarrow 2$)}\newline
$P\not= 0$ and
$\tau^\eta(\rvx \perp \rvy_1 | \rvy_2,\rve)$
and
$\tau^\eta(\rvx \perp \rvy_2 | \rvy_1,\rve)
\Rarrow
\tau^\eta(\rvx \perp \rvy_1,\rvy_2 | \rve)$

\end{enumerate}
The function $\tau^\eta:\cali(\rvxdot)\rarrow Bool$
remains to be specified.
$\rvx,\rvy_1,
\rvy_2,\rve$ stand for
mutually exclusive n-tuples of the form
$\rvxvec{K}$ for some $K\in\zn$.
Rules $(a)$ and $(b)$
perform a ``reduction" whereas $(c)$ and $(d)$
perform a ``combination".

An independency
$I=(\cdot \perp \cdot | \cdot)$
has 3 slots. In the above rule
statements, we've denoted all random
variables in the second slot (slot-2) by
the letter $y$ with a subscript.

The above rule statements start
with the rule name, in parenthesis.
Within the parenthesis, to the left of the slash is the name given
by Judea Pearl in Ref.\cite{Pea1}. To the right of the
slash is a new name, first given
in this paper.
In the new rule names,
the symbol $\rarrow$ stands for implication,
and there is one number, indicating
the number of $y$'s
in slot-2, for each independency.
For example,
in rule $1,1'\rarrow 2$,
there are:
one
$y$ in slot-2 of the
first independency,
one $y$ in slot-2 of the second independency,
two $y$'s in slot-2 of the third independency.
The prime in $1'$ indicates that, besides there being
one $y$ in slot-2, there also is one $y$
in slot-3.

Note that in rule (d) above, we specify that $P\neq 0$.
That's because, as we shall see, this rule
arises from one of those unusual cases,
mentioned earlier,
in which dividing by a probability
causes trouble.
Later on, we will state and prove
theorems whose proof assumes rule (d).
The fact that such theorems assume rule (d)
will show up in that they inherit $P\neq 0$
as one of their premises.

Next we will show that the reduction and
combination rules are obeyed by
$\tau^A$ and $\tau^{CMI}$.

\begin{theo}\label{eq-class-irules}
The above reduction and combination rules are true
in classical physics with $\eta=P$.
\end{theo}
\proof The classical CMI satisfies

\beq
\overbrace{H(\rvx:\rvy_1,\rvy_2|\rve)}^{h_1}=
\overbrace{H(\rvx:\rvy_1|\rvy_2,\rve)}^{h_2}+
\overbrace{H(\rvx:\rvy_2|\rve)}^{h_3}
\;.
\eeq
Permuting $y_1$ and $y_2$ in the
previous equation
yields

\beq
\overbrace{H(\rvx:\rvy_2,\rvy_1|\rve)}^{h_4}=
\overbrace{H(\rvx:\rvy_2|\rvy_1,\rve)}^{h_5}+
\overbrace{H(\rvx:\rvy_1|\rve)}^{h_6}
\;.
\eeq
Recall that the CMI is non-negative.
\begin{itemize}
\item proof of (a)($2\rarrow 1$):$h_1=0\Rarrow h_3=0$.
\item proof of (b)($2\rarrow 1'$):$h_1=0\Rarrow h_2=0$.
\item proof of (c)($1',1\rarrow 2$):$h_2=h_3=0\Rarrow h_1=0$.
\item proof of (d)($1',1'\rarrow 2$):
We want to show that
$h_2=h_5=0{\Rarrow} h_1=0$. Why would this be?
$h_2=0$ and $h_5=0$ imply, respectively,
\bsub
\beq
P(x,y_1,y_2,e) = P(x|y_2,e)P(y_1|y_2,e)P(y_2,e)
\;,
\label{eq-prob-x-y1-y2-e}
\eeq
and

\beq
P(x,y_1,y_2,e) = P(x|y_1,e)P(y_2|y_1,e)P(y_1,e)
\;.
\label{eq-prob-x-y2-y1-e}
\eeq
\esub
We can equate the right hand sides of the
two previous equations, and then divide both
sides of the resulting equation by $P(y_1,y_2,e)$
(here we use $P\neq 0$).
This yields (i) below.
Since we can vary $y_1$ and $y_2$ independently
in equation (i) below,
equation (ii) follows.

\beq
P(x|y_1,e)
\stackrel{(i)}{=} P(x|y_2,e)
\stackrel{(ii)}{=}P(x|e)
\;.
\label{eq-x-indep-y1-y2}
\eeq
Combining Eqs.(\ref{eq-prob-x-y1-y2-e}) and (\ref{eq-x-indep-y1-y2})
then yields,

\beq
P(x,y_1,y_2, e)
=P(x|e)P(y_1,y_2,e)
\;,
\eeq
which, in turn, yields

\beq
P(x,y_1,y_2| e)
=P(x|e)P(y_1,y_2|e)
\;.
\eeq
\end{itemize}
\mbox{}\qed

\begin{theo}
The above reduction and combination rules are true
in quantum physics with $\eta=A$.
\end{theo}
\proof
\begin{itemize}
\item proof of (a)($2\rarrow 1$): The
premise is that

\beq
A(x,y_1,y_2|e) = A(x|e) A(y_1,y_2|e)
\;.
\label{eq-type-a-rule-a-probs}
\eeq
Eq.(\ref{eq-type-a-rule-a-probs})
implies

\beq
P(x,y_1,y_2|e) = P(x|e) P(y_1,y_2|e)
\;.
\eeq
Summing both sides of the previous
equation over $y_1$ yields

\beq
P(x,y_2|e) = P(x|e) P(y_2|e)
\;.
\label{eq-prob-x-y2-at-e}
\eeq
Eq.(\ref{eq-type-a-rule-a-probs}) also
implies

\beq
\theta(x,y_1,y_2|e) = \theta(x|e) \theta(y_1,y_2|e)
\;.
\label{eq-type-a-rule-a-theta}
\eeq
If, in
the previous equation, we
set $y_1$ to its
reference state $y^o_1$, we get

\beq
\theta(x,y_2|e) = \theta(x|e) \theta(y_2|e)
\;.
\label{eq-theta-x-y2-at-e}
\eeq
Combining Eqs.(\ref{eq-prob-x-y2-at-e})
and (\ref{eq-theta-x-y2-at-e}) yields

\beq
A(x,y_2|e) = A(x|e) A(y_2|e)
\;.
\eeq

\item proof of (b)($2\rarrow 1'$): One has

\beq
A(x|y_1,y_2,e)
\stackrel{(i)}{=}
A(x|e)
\stackrel{(ii)}{=}
A(x|y_2,e)
\;.
\eeq
(i)
follows from the premise
 $\tau^A(\rvx \perp \rvy_1,\rvy_2 | \rve)$.
Plugging the premise into rule (a)$(2\rarrow 1)$
gives (ii).

\item proof of (c)($1',1\rarrow 2$): One has

\beq
A(x|y_1,y_2,e)
\stackrel{(i)}{=}
A(x|y_2,e)
\stackrel{(ii)}{=}
A(x|e)
\;.
\eeq
(i)
follows from the part
$\tau^A(\rvx \perp \rvy_1 |\rvy_2, \rve)$
of the premise.
(ii)
follows from the other part
$\tau^A(\rvx \perp \rvy_2 | \rve)$
of the premise.
\item proof of (d)($1',1'\rarrow 2$):
The premise is that
\beq
A(x|y_1,y_2,e) = A(x|y_2,e)
\;,
\label{eq-type-a-rule-b-premise-1}
\eeq
and

\beq
A(x|y_1,y_2,e) = A(x|y_1,e)
\;.
\eeq
Thus, if $A(x,y_1,y_2,e)\neq 0$,

\beq
A(x|y_2,e)=A(x|y_1,e)=A(x|e)
\;.
\label{eq-type-a-rule-b-compare}
\eeq
Combining Eqs.(\ref{eq-type-a-rule-b-premise-1})
and (\ref{eq-type-a-rule-b-compare}) now yields

\beq
A(x|y_1,y_2,e) = A(x|e)
\;.
\eeq
\end{itemize}
\mbox{}\qed

Exercise for reader: Find out whether
$\tausep{},\tau^{CMI}$ and $\tau^{CMI'}$
satisfy the reduction and combination rules.

\subsection{Probabilistic I-sets}
\label{sec-prob-indep}

In this section, we define
certain {\bf probabilistic
I-sets}; that is, I-sets whose
members are defined in terms of
a probability distribution (or a meta
density matrix).

First consider classical physics.
For any $P\in pd(St_{\rvxdot})$, define

\beq
\cali(P) = \{ I:
I=(\rvxvec{J} \perp \rvxvec{K} | \rvxvec{E})
;J,K,E\subset \zn\;{\rm \;are\;disjoint}
;\tau^P(I)\}
\;.
\eeq

Next consider quantum physics. For
any meta density matrix $\mu\in dm(\calh_\rvxdot)$
of the form
$\mu = \proj(\sum_{x.} A(x.)\ket{x.})$, let
\beq
\cali(A) = \{ I:
I=(\rvxvec{J} \perp \rvxvec{K} | \rvxvec{E})
;J,K,E\subset \zn\;{\rm \;are\;disjoint}
;\tau^A(I)\}
\;.
\eeq

For $\eta=P,A$, when we say that an {\bf I-set $\cali$
is satisfied by $\eta$},
we will mean that $\tau^\eta(I)$
for all $I\in \cali$ (or, equivalently,
$\cali\subset \cali(\eta)$).

\section{Bayesian Networks}
In this section, we show that
any probability distribution
can be represented by a fully connected
DAG. We also show that
any quantum density matrix can be
represented by a fully connected DAG.
In classical and quantum physics, omitting
certain arrows from this fully connected graph
 indicates certain
probabilistic independencies.

\subsection{Chain Rule and \\
Factorization According to a Graph}
In this section, we define a
chain rule and factorization according to a
DAG, both for classical and quantum physics.

First consider classical physics. Let
$P\in dm(\calh_\rvxdot)$. For $N=3$,
the $P$ chain rule is

\beq
\underbrace{P(x_3, x_2, x_1)}_7 =
\underbrace{P(x_3|x_2, x_1)}_4
\underbrace{P(x_2|x_1)}_2
\underbrace{P(x_1)}_1
\;.
\label{eq-cla-chain-rule-n3}
\eeq
We have indicated under each
conditional probability the number
of degrees of freedom that it
holds, assuming that $x_1,x_2,x_3\in Bool$.
For arbitrary $N$, the
$P$ chain rule is

\beq
P(x.) = \prod_{j=1}^N P(x_j|\xvec{Z_{1,j-1}})
\;.
\label{eq-cla-chain-rule}
\eeq

Now consider quantum physics.
Suppose $\mu\in dm(\calh_\rvxdot)$
is a meta density matrix of the form
$\mu = \proj(\sum_{x.} A(x.)\ket{x.})$.
In analogy to Eq.(\ref{eq-cla-chain-rule-n3}),
 we would like the
$A$ chain rule for $N=3$ to be

\beq
A(x_3,x_2,x_1) = A(x_3|x_2,x_1) A(x_2|x_1) A(x_1)
\;.
\label{eq-qua-chain-rule-n3}
\eeq

The $P$ chain rule
 Eq.(\ref{eq-cla-chain-rule-n3})
 was stated
without proof, because the equation is
well known, and very easy to
prove. On the other hand,
the $A$ chain rule
Eq.(\ref{eq-qua-chain-rule-n3})
is new,
so we prove it next.

From various definitions
in
Section \ref{sec-cond-amp}, we get

\bsub
\label{eq-3-cond-theta}
\beq
\theta(x_3|x_2,x_1) =
\theta(x_3,x_2,x_1)- \theta(x^o_3,x_2,x_1)
\;,
\eeq

\beq
\theta(x_2|x_1) =
\theta(x^o_3,x_2,x_1)- \theta(x^o_3,x^o_2,x_1)
\;,
\eeq
and

\beq
\theta(x_1) =
\theta(x^o_3,x^o_2,x_1)- \theta(x_3^o,x^o_2,x^o_1)
\;.
\eeq
\esub
Summing Eqs.(\ref{eq-3-cond-theta})
(more precisely, equating the sum of the left
hand sides of Eqs.(\ref{eq-3-cond-theta})
to the sum of the right hand sides) yields

\beq
\theta(x_3|x_2,x_1)+
\theta(x_2|x_1)+
\theta(x_1)=
\theta(x_3,x_2,x_1)
\;.
\label{eq-qua-theta-rule-n3}
\eeq
The previous equation, and
the $P$ chain rule,
together imply:

\beq
\underbrace{e^{i\theta(x_3,x_2,x_1)}}_{7}
\underbrace{\rootp(x_3,x_2,x_1)}_7
=
\underbrace{e^{i\theta(x_3|x_2,x_1)}}_{4}
\underbrace{\rootp(x_3|x_2,x_1)}_4
\underbrace{e^{i\theta(x_2|x_1)}}_{2}
\underbrace{\rootp(x_2|x_1)}_2
\underbrace{e^{i\theta(x_1)}}_{1}
\underbrace{\rootp(x_1)}_1
\;.
\eeq
We have indicated under each quantity
the number of degrees of freedom it holds,
assuming $x_1,x_2,x_3\in Bool$.
The previous equation is equivalent to
Eq.(\ref{eq-qua-chain-rule-n3}), which
we set out to prove.
For arbitrary $N$,
Eq.(\ref{eq-qua-theta-rule-n3})
generalizes to

\beq
\theta(x.) = \sum_{j=1}^N
\theta(x_j | \xvec{Z_{1,j-1}})
\;.
\eeq
The previous equation, and
Eq.(\ref{eq-cla-chain-rule})
(the $P$ chain rule), together
imply the $A$ chain rule:

\beq
A(x.) = \prod_{j=1}^N A(x_j|\xvec{Z_{1,j-1}})
\;.
\label{eq-qua-chain-rule}
\eeq

Note that the conditional amplitudes
$A(x_j|\xvec{Z_{1,j-1}})$
used above have {\bf constrained phases (CP)},
meaning that their
phases are subject to the constraint
that $A(x_j^o|\xvec{Z_{1,j-1}})$
be real for all $\xvec{Z_{1,j-1}}$.
Let $M$ be
the matrix with
entries $A(x_j|\xvec{Z_{1,j-1}})$,
with the rows of $M$ labelled by
the states of $\rvx_j$
and the columns labelled by the
states of $\rvxvec{Z_{1,j-1}}$.
CP means that
$M$ must have one row (the one with $x_j=x_j^o$)
consisting entirely of real numbers.
On the other hand, Quantum Fog allows
conditional
amplitudes $A(x_j|\xvec{Z_{1,j-1}})$
with {\bf free phases (FP)}, meaning
that the phases of $A(x_j|\xvec{Z_{1,j-1}})$
 are arbitrary.
Clearly, it is often
convenient, not just in Quantum Fog,
to allow FP amplitudes.
Luckily, one can always replace an
FP amplitude
$A(x_j|\xvec{Z_{1,j-1}})$
by a product of CP
amplitudes.
This is how. To simplify our notation,
let $x_j\rarrow a$
and $\xvec{Z_{1,j-1}}\rarrow b$.
Replace an
FP amplitude $A(b|a)$
by a product of three CP amplitudes
$A(b|a''),
A(a''|a')$
and
$A(a'|a)$:

\beq
A(b|a)=
\sum_{a',a''}
A(b|a'')
A(a''|a')
A(a'|a)
\;,
\label{eq-lp-equals-3cp}
\eeq
where $a',a''\in St_\rva$.
$A(b|a)$ can be interpreted as the
transition matrix of node $\rvb$ in
a subgraph
\beq(\rvb)\larrow(\rva)\;.\eeq
This subgraph is being replaced by
a Markov-chain graph
\beq(\rvb)\larrow(\rva'')\larrow(\rva')\larrow(\rva)\;.\eeq
Define the following
matrices:

\beq
[A(b|a)] = F
\;,\;\;
[A(b|a'')] = C_1
\;,\;\;
[A(a''|a')] = C_2
\;,\;\;
[A(a'|a)] = C_3
\;.
\eeq
Eq.(\ref{eq-lp-equals-3cp}),
expressed in matrix form, is

\beq
F = C_1 C_2 C_3
\;.
\eeq
Suppose the first row of $F$ is
$[x_1e^{i\phi_1},x_2e^{i\phi_2},\ldots, x_{N_\rva}e^{i\phi_{N_\rva}}]$,
where $x_j,\phi_j\in\RR$.
Let

\beq
\begin{array}{l}
C_1=
F\diag(e^{-i\phi_1},e^{-i\phi_2},,e^{-i\phi_3},\ldots, e^{-i\phi_{N_\rva}})
\\
C_2=
\diag(1,e^{i\phi_2},e^{i\phi_3},\ldots, e^{i\phi_{N_\rva}})
\\
C_3=
\diag(e^{i\phi_1},1,1,\ldots, 1)
\end{array}
\;.
\eeq
The matrices $C_1,C_2,C_3$ all have
at least one row that consisting entirely of reals,
so these matrices specify CP
amplitudes. (If global phases are allowed,
only 2 C's are necessary).

We end this section by
defining graphic
factorization.
In classical physics, we
say $P\in pd(St_\rvxdot)$ {\bf factors
according to}  $G\in DAG(\rvxdot)$ iff

\beq
P(x.) = \prod_{j=1}^N P(x_j|\xvec{pa(j)})
\;.
\eeq
In quantum physics, for
a meta density matrix
$\mu\in dm(\calh_\rvxdot)$
of the form
$\mu = \proj(\sum_{x.} A(x.)\ket{x.})$,
we say $A$
{\bf factors according to}  $G\in DAG(\rvxdot)$
 iff

\beq
A(x.) = \prod_{j=1}^N A(x_j|\xvec{pa(j)})
\;.
\eeq

By virtue of the
$P$ (ditto, $A$)
chain rule,
any probability distribution (ditto,
probability amplitude) of $\rvxdot$
factors according
to an $N$-node fully-connected
DAG. If the probability distribution (ditto,
probability amplitude) has higher symmetry,
then it may also factor according to another
$N$-node graph that possess
fewer arrows than the fully-connected one.

\subsection{Graphic I-sets}
\label{sec-bnet-graph-iset}

In Section \ref{sec-prob-indep},
we defined some probabilistic I-sets.
The elements of a {\bf probabilistic I-set}
are defined in terms of a probability
distribution (or a meta density matrix).
In this section, we define some
graphic
I-sets for a DAG.
The elements of a {\bf graphic I-set}
are defined with
respect to a graph.

For $G\in DAG(\rvxdot)$, we define (loc=local,
glo=global)

\beq
\cali_{loc}(G)=
\{I: I = (\rvx_j \perp \rvxvec{\neg de(j)}| \rvxvec{pa(j)}),
j\in \zn \}
\;,
\eeq
and

\beq
\cali_{glo}(G)=
\{I: I = (\rvxvec{J} \perp \rvxvec{K}| \rvxvec{E})
;J,K,E\subset \zn\;{\rm \;are\;disjoint}
;\tausep{}(I)\}
\;.
\eeq
The function $\tausep{}:\cali(\rvxdot)\rarrow Bool$
will be defined later on.

For example, if $G$ is the graph of
Fig.\ref{fig-diamond-graphs}(a),
then

\beq
\cali_{loc}(G)=\{
(\rvx_3 \perp \rvx_2 | \rvx_1),
(\rvx_4 \perp \rvx_1 | \rvx_2, \rvx_3)
\}
\;.
\eeq

\subsection{Graphic Factorization iff an I-set is satisfied}
In this section, we show
that a probability
distribution (ditto, probability amplitude)
factors according to a DAG iff the
probability distribution (ditto,
probability amplitude) satisfies a graphic I-set.

As motivation for the main theorem of this
section, let $G$ be the DAG of
Fig.\ref{fig-diamond-graphs}(a).
Note that $\cali_{loc}(G)\subset \cali(P)$
iff $\tau^P(I)$ for all $I\in\cali_{loc}(G)$.
Therefore, for the graph $G$ of
Fig.\ref{fig-diamond-graphs}(a),
$\cali_{loc}(G)\subset \cali(P)$
is equivalent to

\beq
\left\{
\begin{array}{l}
\tau^P(\rvx_3 \perp \rvx_2 | \rvx_1)\\
\tau^P(\rvx_4 \perp \rvx_1 | \rvx_2, \rvx_3)
\end{array}
\right.
\;.
\label{eq-I-set-diamond-graphs}
\eeq
Eq.(\ref{eq-I-set-diamond-graphs}) is
itself equivalent to

\beq
\left\{
\begin{array}{l}
P(x_3|x_2,x_1)=P(x_3|x_1)\\
P(x_4|x_3, x_2,x_1)=P(x_4|x_3,x_2)
\end{array}
\right.
\;.
\label{eq-p-reduct-diamond}
\eeq
Define $P_{chain}$ and $P_{graph}$
as the following two probability distributions
of $\rvxdot$:

\beq
P_{chain}(x.)=
P(x_4|x_3,x_2,x_1)
P(x_3|x_2,x_1)
P(x_2|x_1)
P(x_1)
\;,
\label{eq-p-chain-diamond}
\eeq
and

\beq
P_{graph}(x.)=
P(x_4|x_3,x_2)
P(x_3|x_1)
P(x_2|x_1)
P(x_1)
\;.
\label{eq-p-graph-diamond}
\eeq
$P_{chain}$ comes from
the $P$ chain rule and
$P_{graph}$ from the definition
of factorization according to the graph
of Fig.\ref{fig-diamond-graphs}(a).
From Eqs.(\ref{eq-p-reduct-diamond}),
(\ref{eq-p-chain-diamond}) and
(\ref{eq-p-graph-diamond}),
it is clear that:
If $\tau^P(I)$ for all $I\in\cali_{loc}(G)$,
 then $P_{chain}=P_{graph}$.
The converse statement is also true.
This is why.
$P_{chain}=P_{graph}$
implies

\beq
P(x_4|x_3,x_2,x_1)
P(x_3|x_2,x_1)
=
P(x_4|x_3,x_2)
P(x_3|x_1)
\;.
\label{eq-pchain-pgraph-diamond}
\eeq
Summing both sides over $x_4$ gives
$P(x_3|x_2,x_1)=P(x_3|x_1)$.
Combining this result with
Eq.(\ref{eq-pchain-pgraph-diamond}) then gives
$P(x_4|x_3,x_2,x_1)=P(x_4|x_3,x_2)$.
Thus, Eqs.(\ref{eq-p-reduct-diamond}) are
obeyed.
We have just proven, albeit
only for the graph of
Fig.\ref{fig-diamond-graphs}(a),
the following theorem:

\begin{theo}
Suppose $G\in DAG(\rvxdot)$ and
$P\in pd(St_\rvxdot)$.\newline
$P$ factors according to $G$
iff
$\cali_{loc}(G)\subset \cali(P)$.
\end{theo}
\proof The proof is a special case
of the proof of the next theorem.
\qed

\begin{theo}
Suppose $G\in DAG(\rvxdot)$ and
$\mu\in dm(\calh_\rvxdot)$ is a
meta density matrix of the form
$\mu = \proj(\sum_{x.} A(x.)\ket{x.})$.
\newline
$A$ factors according to $G$
iff
$\cali_{loc}(G)\subset \cali(A)$.
\end{theo}
\proof
Without loss of generality, we can assume that
the nodes are labelled so that
$pa(j)\subset Z_{1,j-1}$ for all $j$.
This means that we can always add arrows to $G$
until we generate a fully connected graph
$\overline{G}$ such that
$pa(j)= Z_{1,j-1}$. We will call
$\overline{G}$
a {\bf proper fully-connected extension} of $G$.
What we need to prove can now be
rephrased as:
\beq
A_{chain} = A_{graph}
\;\;{\rm iff}\;\;
\tau^A(\rvx_j \perp \rvxvec{\neg de(j)}| \xvec{pa(j)})
\;\;\forall j
\;,
\eeq
where

\beq
A_{chain}(x.) =
\prod_{j=1}^N A(x_j|\xvec{Z_{1,j-1}})
\;,\;\;
A_{graph}(x.) =
\prod_{j=1}^N A(x_j|\xvec{pa(j)})
\;.
\eeq

($\Larrow$) Define $Z'_{1,j-1} = Z_{1,j-1} -pa(j)$.
Since $\tau^A(\rvx_j \perp \rvxvec{\neg de(j)}| \xvec{pa(j)})$
and $Z'_{1,j-1}\subset Z_{1,j-1}\subset \neg de(j)$,
it follows from reduction rule $2\rarrow 1$ that
$\tau^A(\rvx_j \perp \rvxvec{Z'_{1,j-1}}| \xvec{pa(j)})$.
Thus,
\bsub
\beqa
A(x_j|\xvec{Z_{1,j-1}}) &=&
A(x_j| \xvec{Z'_{1,j-1}},\xvec{pa(j)})
\\
&=&
A(x_j|\xvec{pa(j)})
\;.
\eeqa
\esub

($\Rarrow$)$A_{chain}=A_{graph}$ implies that

\beq
\prod_{j=1}^N P(x_j|\xvec{Z_{1,j-1}})
=
\prod_{j=1}^N P(x_j|\xvec{pa(j)})
\;.
\label{eq-pchain-is-pgraph}
\eeq

\begin{itemize}
\item
Sum both sides of Eq.(\ref{eq-pchain-is-pgraph})
over $\xvec{Z_{2,N}}$. Get
$P(x_1)=P(x_1)$.
\item
Divide both sides of
Eq.(\ref{eq-pchain-is-pgraph}) by $P(x_1)$, and then
sum both sides over $\xvec{Z_{3,N}}$. Get
$P(x_2|x_1)=P(x_2|\xvec{pa(2)})$.
\item
Divide both sides of
Eq.(\ref{eq-pchain-is-pgraph}) by $P(x_2,x_1)$, and then
sum both sides over $\xvec{Z_{4,N}}$. Get
$P(x_3|x_2, x_1)=P(x_3|\xvec{pa(3)})$.
\item
Divide both sides of
Eq.(\ref{eq-pchain-is-pgraph}) by $P(x_3,x_2,x_1)$, and then
sum both sides over $\xvec{Z_{5,N}}$. Get
$P(x_4|x_3,x_2,x_1)=P(x_4|\xvec{pa(4)})$.
\item
And so on.
\end{itemize}
Thus, by induction,

\beq
P(x_j | \xvec{Z_{1,j-1}})=
P(x_j | \xvec{pa(j)})
\;
\label{eq-prob-prev-to-pa}
\eeq
for all $j$.

$A_{chain}=A_{graph}$ also implies that

\beq
\sum_{j=1}^N\theta(x_j|\xvec{Z_{1,j-1}})=
\sum_{j=1}^N \theta(x_j | \xvec{pa(j)})
\;.
\label{eq-thetachain-is-thetagraph}
\eeq
Recall that
$\theta(x_j^o|\xvec{Z_{1,j-1}}) = 0$.

\begin{itemize}
\item
Set $\xvec{Z_{2,N}}\rarrow \xovec{Z_{2,N}}$
in  Eq.(\ref{eq-thetachain-is-thetagraph}). Get
$\theta(x_1)=\theta(x_1)$.
\item
Set $\xvec{Z_{3,N}}\rarrow \xovec{Z_{3,N}}$
in  Eq.(\ref{eq-thetachain-is-thetagraph}) and subtract
$\theta(x_1)$ from both sides. Get
$\theta(x_2|x_1)=\theta(x_2|\xvec{pa(2)})$.
\item
Set $\xvec{Z_{4,N}}\rarrow \xovec{Z_{4,N}}$
in  Eq.(\ref{eq-thetachain-is-thetagraph}) and subtract
$\theta(x_2,x_1)$ from both sides. Get
$\theta(x_3|x_2, x_1)=\theta(x_3|\xvec{pa(3)})$.
\item
Set $\xvec{Z_{5,N}}\rarrow \xovec{Z_{5,N}}$
in  Eq.(\ref{eq-thetachain-is-thetagraph}) and subtract
$\theta(x_3, x_2,x_1)$ from both sides. Get
$\theta(x_4|x_3, x_2, x_1)=\theta(x_4|\xvec{pa(4)})$.
\item
And so on.
\end{itemize}
Thus, by induction,

\beq
\theta(x_j | \xvec{Z_{1,j-1}})=
\theta(x_j | \xvec{pa(j)})
\;
\label{eq-theta-prev-to-pa}
\eeq
for all $j$.
Combining Eq.(\ref{eq-prob-prev-to-pa})
and (\ref{eq-theta-prev-to-pa}) yields

\beq
A(x_j | \xvec{Z_{1,j-1}})=
A(x_j | \xvec{pa(j)})
\;
\label{eq-amp-prev-to-pa}
\eeq
for all $j$. Note that
$\neg de(j)\supset Z_{1,j-1}$.
From a proper fully-connected extension of
$G$, it is clear that

\beq
A(x_j | \xvec{\neg de(j)})=
A(x_j | \xvec{Z_{1,j-1}})
\;
\eeq
for all $j$. Combining
the previous two equations yields

\beq
A(x_j | \xvec{\neg de(j)})=
A(x_j | \xvec{pa(j)})
\;
\eeq
for all $j$.
Define $\neg de'(j) = \neg de(j)-pa(j)$.
The previous equation can be
written as $A(x_j | \xvec{\neg de'(j)}, \xvec{pa(j)})=
A(x_j | \xvec{pa(j)})$,
which means that
$\tau^A(\rvx_j\perp \rvxvec{\neg de'(j)}| \rvxvec{pa(j)})$.
Thus,
$\tau^A(\rvx_j\perp \rvxvec{\neg de(j)}| \rvxvec{pa(j)})$.
\qed
\subsection{Going Global}

In the last section, we showed
that a probability distribution $P$
(or a probability amplitude $A$) factors according
to a DAG iff it satisfies a
certain non-global, graphic I-set.
Does a similar result hold
if the non-global
 graphic I-set
 is replaced by a global graphic one?
This section will be devoted to
answering this question.

\newpage
\begin{figure}[h]
    \begin{center}
    \epsfig{file=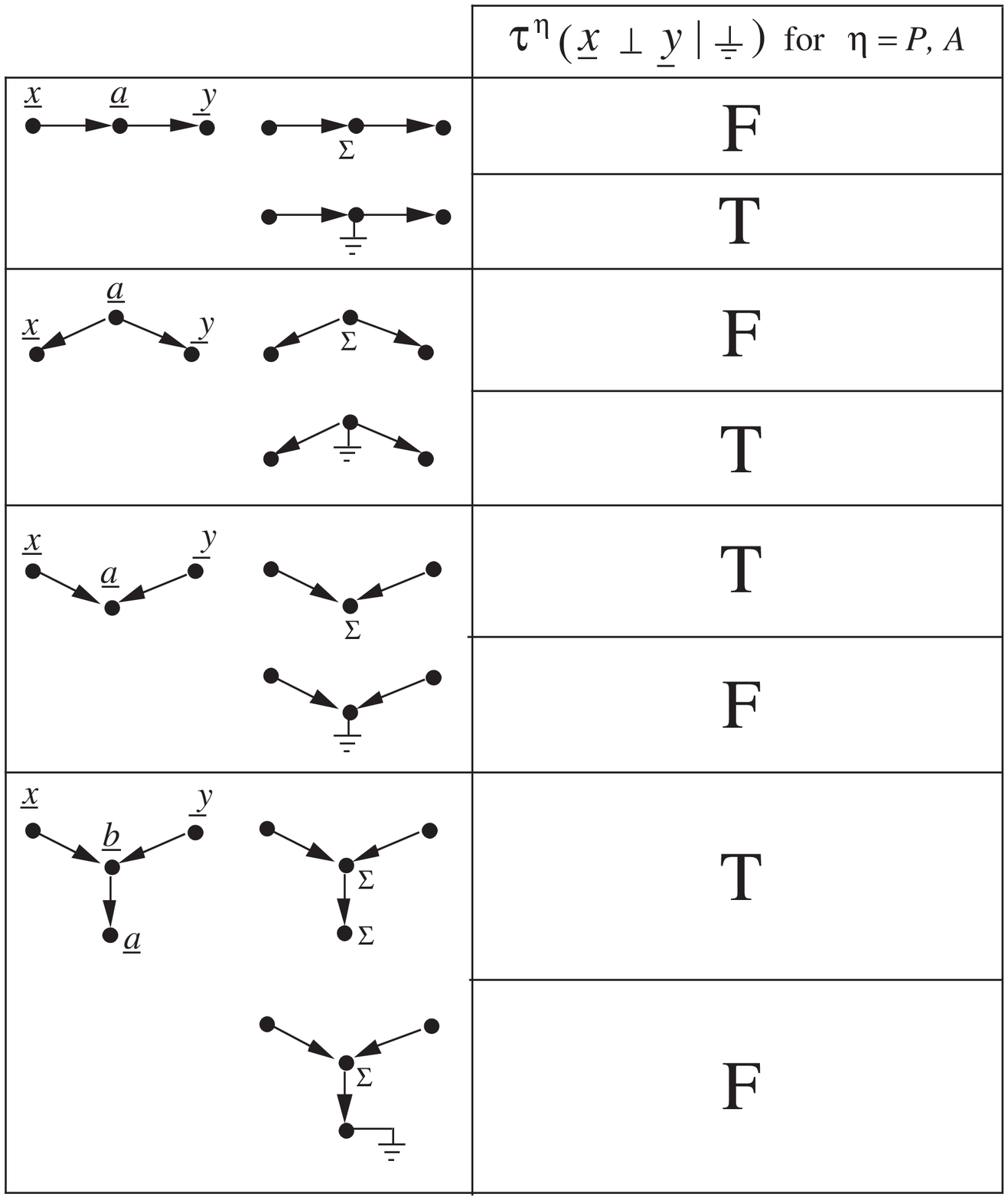, width=4in}
    \caption{Some simple Bayesian nets and
   their truth value for
    $\tau^P(\rvx \perp \rvy| \ground)$,
    which is the same as their truth value for
    $\tau^A(\rvx \perp \rvy| \ground)$.
    The independency $(\rvx \perp \rvy| \ground)$
    is conditioned on the
    grounded nodes. T=true, F=false.}
    \label{fig-dsep-main-cases}
    \end{center}
\end{figure}

To develop some intuition, we begin
by considering
Fig.\ref{fig-dsep-main-cases},
which
shows some simple Bayesian net examples.

Column 1 of Fig.\ref{fig-dsep-main-cases}
 shows
four DAGs in which, respectively,
node $\rva$ is:

\begin{enumerate}
\item
a serial node of
a path from $\rvx$ to $\rvy$,
\item
(``common cause" graph) a divergence node of
a path from $\rvx$ to $\rvy$
\item
(``common effect", ``explaining away" graph) a convergence
(a.k.a. collider) node of a path from $\rvx$ to $\rvy$
\item
the descendant of a collider node of
a path from $\rvx$ to $\rvy$.
\end{enumerate}

Column 2
of Fig.\ref{fig-dsep-main-cases}
illustrates two special cases
of the graphs in column 1:
(1) no node is grounded,
(2) only node $\rva$ is grounded.
Nodes decorated with a $\Sigma$
are summed over for $\eta=P$ and
traced over for $\eta=A$.

In  Fig.\ref{fig-dsep-main-cases},
the argument of $\tau^\eta$ is an
independency whose
third slot
  is filled with
the grounded nodes. In Fig.\ref{fig-dsep-main-cases},
the grounded nodes are always
either $\rva$ or nothing. An independency
with no grounded nodes is unconditional.
In the classical physics case,
column 3 of
Fig.\ref{fig-dsep-main-cases}
gives the truth values (T=true, F=false)
of $\tau^P(I)$,
for the graphs in column 2.
In the quantum physics case,
column 3
gives the values of
$\tau^A(I)$.

Next, we show how we calculated the
truth values of $\tau^A(I)$ in
Fig.\ref{fig-dsep-main-cases}.
Let $A=\rootp e^{i\theta}$.

Rows 1 and 2 (graphs in which $\rva$
is a serial node) satisfy

\beq
A(x,y,a) = A(y|a)A(a|x)A(x)=A(y|a)A(x|a)A(a)
\;,
\label{eq-dsep-table-serial}
\eeq
for all $x,y,a$.
Eq.(\ref{eq-dsep-table-serial})
implies

\beq
P(x,y)= \sum_a P(y|a)P(x|a)P(a)\neq P(x)P(y)
\;,\;\; A(x,y)\neq A(x)A(y)
\;.
\eeq
$\tau^P(\rvx\perp\rvy)$ is false
so $\tau^A(\rvx\perp\rvy)$ is false too.
Eq.(\ref{eq-dsep-table-serial})
implies

\beq
A(x,y|a)=A(y|a)A(x|a)
\;,
\eeq
so $\tau^A(\rvx\perp\rvy|\rva)$ is true.

Rows 3 and 4  with divergence node $\rva$
must have the same
truth values  as rows 1 and 2 with serial
node $\rva$. That's because
the Bayesian nets
\beq(\rvx)\rarrow(\rva)\rarrow(\rvy)\eeq
and
\beq(\rvx)\larrow(\rva)\rarrow(\rvy)\eeq
are indistinguishable: they both
represent the same full joint amplitude.
Indeed,
$A_1(x,y,a) = A(y|a)A(a|x)A(x)$
for the first and
$A_2(x,y,a) = A(y|a)A(x|a)A(a)$
for the second, and $A_1 = A_2$.
\footnote{
Sometimes, some of the arrows of a
classical Bayesian can be reversed
without changing the
full joint
probability distribution of the net.
General rules have been given in the literature
(see Ref.\cite{Kol}) for deciding
which arrows can be reversed with
impunity.
Similar rules apply for
quantum Bayesian nets.
}

Rows 5 and 6 (graphs in which $\rva$
is a collider node) satisfy

\beq
A(x,y,a) = A(a|x,y)A(y)A(x)
\;,
\eeq
for all $x,y,a$. Therefore,
$P(x,y) = P(x)P(y)$,
$\theta(x,y) = \theta(x)\theta(y)$.
Hence, $\tau^A(\rvx\perp\rvy)$ is true.
$\tau^P(\rvx\perp\rvy|\rva)$ is false
so $\tau^A(\rvx\perp\rvy|\rva)$ is false too.

Rows 7 and 8 (graphs in which $\rva$
is a descendant of a collider node) satisfy
\beq
A(x,y,a,b)=A(a|b)A(b|x,y)A(x)A(y)
\;,
\eeq
for all $x,y,a,b$.
Therefore,
$P(x,y) = P(x)P(y)$,
$\theta(x,y) = \theta(x)\theta(y)$.
Hence, $\tau^A(\rvx\perp\rvy)$ is true.
$\tau^P(\rvx\perp\rvy|\rva)$ is false
so $\tau^A(\rvx\perp\rvy|\rva)$ is false too.

Note that the calculations of the
truth values of $\tau^P(I)$ in
Fig.\ref{fig-dsep-main-cases}
are a special case of the just
presented calculations of the
truth values of $\tau^A(I)$.

The moral of Fig.\ref{fig-dsep-main-cases},
is that grounding
a serial node or a divergence node
interrupts information
transmission
between $\rvx$ and $\rvy$.
A non-vacuous message has variation in it,
and a grounded node in its path prevents
transmission of this variation.
However, grounding a collider or
a descendant of a collider
has the opposite effect: it allows
information transmission (this is called
the ``explaining away" phenomenon).

So far,
this section has presented
merely  anecdotal evidence.
Next, we will state and prove
some general theorems.

Consider any $G\in DAG(\rvxdot)$.
Suppose $J,K,E\subset \zn$ are disjoint sets.
Let $I = (\rvxvec{J}\perp \rvxvec{K}|\rvxvec{E})$.
We will abbreviate  ``dependency separation" by ``d-sep"
or just ``sep".
We define the function
$\tausep{}:\cali(\rvxdot)\rarrow Bool$
by the statement:
$\tausep{}(I)$ is true
iff all
paths $\gamma$ in $G$ from a node
in $\rvxvec{J}$ to a node in $\rvxvec{K}$
are blocked by $\rvxvec{E}$.
We
say ``$\gamma$ is blocked by $\rvxvec{E}$" iff
there exists a node $\rvx_i\in\gamma$ that
satisfies one of the following:
\begin{enumerate}
\item $\rvx_i$ is a non-collider of $\gamma$ and
$i\in E$.
\item $\rvx_i$ is a collider of $\gamma$ and
$\overline{de}(i)\cap E=\emptyset$
\end{enumerate}

\begin{theo}\label{th-bnet-prob-dsep}
(Classical d-Separation Theorem)
Suppose $G\in DAG(\rvxdot)$ and
$P\in pd(St_\rvxdot)$.
If $P$ factors according to $G$ then
$\cali_{glo}(G)\subset \cali(P)$.
\end{theo}
\proof
The proof of this theorem can be found
in the literature\cite{Kol}\cite{Lau}.
\qed

\begin{theo}\label{th-bnet-amp-dsep}
(Quantum d-Separation Theorem)
Suppose $G\in DAG(\rvxdot)$ and
$\mu\in dm(\calh_\rvxdot)$ is a
meta density matrix of the form
$\mu = \proj(\sum_{x.} A(x.)\ket{x.})$.
If $A$ factors according to $G$ then
$\cali_{glo}(G)\subset \cali(A)$.
\end{theo}
\proof
The proof of this theorem
is a simple generalization of
the proof of Theorem \ref{th-bnet-prob-dsep}.
\qed

One can also prove a weak converse of
the d-Separation Theorem. The weak converse
theorem\cite{Kol} shows that
$\cali_{glo}(G)$ is
in some sense the maximal set for which
the d-Separation Theorem holds.
For this reason, Ref.\cite{Kol}
describes the d-Separation Theorem
as a proof of soundness and its
weak converse as a proof of completeness.

\section{Markov Networks}
In this section, we show that
any probability distribution
can be represented by a fully connected
UG. We also show that
any quantum density matrix can be
represented by a fully connected UG.
In classical and quantum physics, omitting
certain links from this fully connected graph
 indicates certain
probabilistic independencies.

\subsection{Power-set Rule and \\
Factorization According to a Graph}

In this section, we define a
power-set rule
 and factorization according to an
UG, both for classical and quantum physics.
The $P$ power-set rule is well known, but not
by that name, which is ours.
In some sense, the
$P$ (ditto, $A$) power-set rule is to Markov nets what
the $P$ (ditto, $A$) chain rule is to  Bayesian nets.

\begin{theo}(P Power-set Rule)
Any $P\in pd(St_\rvxdot)$ can be expressed as
\beq
P(x.)=
\prod_{J:J\subset \zn} e^{\lam(\xvec{J})}
\;,
\label{eq-prob-power-set-rule}
\eeq
where $\lam(\xvec{J})$ is defined by

\beq
\lam(\xvec{J}) =
\sum_{J':J'\subset J}(-1)^{|J-J'|}
\ln P(\xvec{J'}, \xovec{(J')^c})
\;.
\label{eq-prob-lam-def}
\eeq
(Note that if for some
point $x.'$, $P(x.')=0$ , then
$\lam(\xpvec{J})=-\infty$ for some $J$.
Instead of permitting
such infinities,
as we do, some authors
restrict this theorem by adding
a premise that $P\neq 0$.)
\end{theo}
\proof The proof is a special case
of the proof of the next theorem.
\qed

\begin{theo}(A Power-set Rule)
Given a meta density matrix $\mu\in dm(\calh_\rvxdot)$
of the form
$\mu = \proj(\sum_{x.} A(x.)\ket{x.})$, $A$
 can be expressed as

\beq
A(x.)=
\prod_{J: J\subset \zn} e^{\lam(\xvec{J})}
\;,
\label{eq-amp-power-set-rule}
\eeq
where $\lam(\xvec{J})$ is defined by

\beq
\lam(\xvec{J}) =
\sum_{J':J'\subset J}(-1)^{|J-J'|}
\ln A(\xvec{J'}, \xovec{(J')^c})
\;.
\label{eq-amp-lam-def}
\eeq
(Note that if for some
point $x.'$, $A(x.')=0$ , then
$Re(\lam(\xpvec{J}))=-\infty$
for some $J$.
Instead of permitting such infinities,
as we do, some authors
restrict this theorem by adding
a premise that $A\neq 0$.)
\end{theo}
\proof
Performing a Mobius Inversion
(see Appendix \ref{app-mobius})
on Eq.(\ref{eq-amp-lam-def}),
we get

\beq
\ln A(\xvec{J}, \xovec{(J)^c})=
\sum_{J':J'\subset J}\lam(\xvec{J'})
\;.
\eeq
Replacing $J$ by $\zn$ in the previous equation
yields:

\beq
\ln A(x.)=
\sum_{J:J\subset \zn}\lam(\xvec{J})
\;.
\eeq
\qed

For $N$ random variables,
the $P$ (ditto, $A$) chain rule
contains $N$ factors whereas
the $P$ (ditto, $A$) power-set rule
contains $2^N$.
Thus, a power-set rule is
not as useful as a chain rule
for practical purposes like numerical calculation.
It is mainly used to  prove
other theorems.

We end this section by
defining graphic
factorization.
In classical physics, we
say $P\in pd(St_\rvxdot)$ {\bf factors
according to}  $G\in UG(\rvxdot)$
iff $P$ can be expressed in the form of

\beq
P(x.)=
\prod_{J\in super-cliques(G)} e^{\lam(\xvec{J})}
\;.
\eeq
In quantum physics, for
a meta density matrix
$\mu\in dm(\calh_\rvxdot)$
of the form
$\mu = \proj(\sum_{x.} A(x.)\ket{x.})$,
we say $A$
{\bf factors according to}  $G\in UG(\rvxdot)$
 iff $A$ can be expressed in the form of
\beq
A(x.)=
\prod_{J\in super-cliques(G)} e^{\lam(\xvec{J})}
\;.
\label{eq-amp-factor-acc-ug}
\eeq

When $G$ is fully connected,
Eq.(\ref{eq-amp-factor-acc-ug}) reduces
to $A(x.)= e^{\lam(x.)}$, which is
always possible.
Thus,
any probability distribution (ditto,
probability amplitude) of $\rvxdot$
factors according
to an $N$-node fully-connected
UG. If the probability distribution (ditto,
probability amplitude) has higher symmetry,
then it may also factor according to another
$N$-node graph that possess
fewer links than the fully-connected one.

\subsection{Graphic I-sets}
In Section \ref{sec-bnet-graph-iset},
we defined some
graphic I-sets for a DAG.
In this section, we define some
graphic
I-sets for an UG.

For $G\in UG(\rvxdot)$, we define (loc=local,
glo=global)

\beq
\cali_{pair}(G)=
\{I: I=(\rvx_{j_1}\perp\rvx_{j_2}|
\rvxvec{\zn-\{j_1,j_2\}}), j_1\not\in ne(j_2)
;  j_1,j_2\in \zn\}
\;,
\eeq

\beq
\cali_{loc}(G)=
\{I: I = (\rvx_j \perp
\rvxvec{\zn - \overline{ne}(j)}| \rvxvec{ne(j)}),
j\in \zn \}
\;,
\eeq
and

\beq
\cali_{glo}G)=
\{I: I = (\rvxvec{J} \perp \rvxvec{K}| \rvxvec{E})
;J,K,E\subset \zn\;{\rm \;are\;disjoint}
;\tausep{}(I)\}
\;.
\eeq
The function $\tausep{}:\cali(\rvxdot)\rarrow Bool$
will be defined later on.

\subsection{Graphic Factorization iff an I-set is satisfied}
In this section, we show
that a probability
distribution (ditto, probability amplitude)
factors according to an UG iff the
probability distribution
(ditto, probability amplitude)
satisfies a graphic I-set.

\begin{theo}\label{th-prob-fact-iff-local-iset}
Suppose $G\in UG(\rvxdot)$ and
$P\in pd(St_\rvxdot)$.
\newline
$P$ factors according to $G$
$\Leftrightarrow$
$\cali_{pair}(G)\subset \cali(P)$.
\end{theo}
\proof The proof is a special case
of the proof of the next theorem.
\qed

\begin{theo}\label{th-amp-fact-iff-local-iset}
Suppose $G\in UG(\rvxdot)$ and
$\mu\in dm(\calh_\rvxdot)$ is a
meta density matrix of the form
$\mu = \proj(\sum_{x.} A(x.)\ket{x.})$.
\newline
$A$ factors according to $G$
$\Leftrightarrow$
$\cali_{pair}(G)\subset \cali(A)$.
\end{theo}
\proof
If the number of nodes $N$ is
one then the theorem is satisfied trivially,
so assume $N\geq 2$.
Recall $A$ factors according to $G$ iff

\beq
A(x.) = \prod_{J\in super-cliques(G)} e^{\lam(\xvec{J})}
\;.
\label{eq-in-proof-amp-factors}
\eeq
Note that $\cali_{pair}(G)\subset \cali(A)$ iff
$\tau^A(\rvx_{j_1} \perp \rvx_{j_2}| \rvxvec{\zn-\{j_1,j_2\}})$
for all $j_1, j_2\in \zn$ such that
$j_1\not\in ne(j_2)$.

($\Larrow$)(This direction would require
a premise $A\neq 0$ if we weren't  permitting
infinite $|\lam(\xvec{J})|$).
Consider any $J\subset \zn$.
Suppose $j_1, j_2$ are any
two elements of $J$ (there
may or may not be a link between
$\rvx_{j_1}$ and $\rvx_{j_2}$
at this point).
Let $\jm$ denote $J-\{j_1,j_2\}$.
Note that for any function $f.:2^J\rarrow \CC$

\beq
\sum_{J':J'\subset J}f_{J'}=
\sum_{\jmp:\jmp\subset \jm}
\left(
f_{\jmp}
+ f_{\jmp\cup\{j_1,j_2\}}
+ f_{\jmp\cup\{j_1\}}
+ f_{\jmp\cup\{j_2\}}
\right)
\;.
\label{eq-jp-to-jpm}
\eeq
Now define $\xi$ by

\beq
\xi = (\xvec{\jmp},\xovec{\zn-\{j_1,j_2\}-\jmp})
\;.
\eeq
Using Eq.(\ref{eq-jp-to-jpm}),
Eq.(\ref{eq-amp-lam-def})
can be re-written as

\beq
\lam(\xvec{J}) =
\sum_{\jmp:\jmp\subset \jm}(-1)^{|\jm-\jmp|}
\ln
\left[
\frac{
A(x_{j_1},x_{j_2}, \xi)A(x^o_{j_1},x^o_{j_2}, \xi)
}{
A(x^o_{j_1},x_{j_2}, \xi)A(x_{j_1},x^o_{j_2}, \xi)
}
\right]
\;.
\label{eq-amp-fancy-lam-def}
\eeq
If $\rvx_{j_1}$ and
$\rvx_{j_2}$ are not in
 the same super-clique,
then there is no link between them.
$j_1\not\in ne(j_2)$ so
$\tau^A(\rvx_{j_1} \perp \rvx_{j_2}| \rvxvec{\zn-\{j_1,j_2\}})$,
so

\beq
A(x'_{j_1},x'_{j_2}|\xi)=
A(x'_{j_1}|\xi)A(x'_{j_2}|\xi)
\;,
\label{eq-xj1-xj2-indep-given-xi}
\eeq
for all
$x'_{j_1}\in N_{\rvx_{j_1}}$
and
$x'_{j_2}\in N_{\rvx_{j_2}}$.
When
 Eq.(\ref{eq-xj1-xj2-indep-given-xi})
 is true,
the right hand side
of Eq.(\ref{eq-amp-fancy-lam-def})
vanishes. In conclusion,
if $j_1, j_2\in J$ but
$\rvx_{j_1}$ and
$\rvx_{j_2}$ are not in
the same super-clique,
then $\lam(\xvec{J})=0.$
In general,
$
A(x.) = \prod_{J:J\subset \zn} e^{\lam(\xvec{J})}
$. (This would require $A\neq 0$ if
infinite $|\lam(\xvec{J})|$ were not permitted.)
But we have shown
that $\lam(\xvec{J})$
vanishes for any $J\subset \zn$
which is not a super-clique of $G$.
Thus, Eq.(\ref{eq-in-proof-amp-factors})
follows.

($\Rarrow$) Let $j_1, j_2\in \zn$
such that $j_1\not\in ne(j_2)$.
Define
$R = \zn - \{j_1,j_2\}$.
$\rvx_{j_1}$ and $\rvx_{j_2}$
must belong to different super-cliques of $G$.
This fact and Eq.(\ref{eq-in-proof-amp-factors})
together imply that
there
exist  sets $R_1, R_2$
(not necessarily disjoint) such that
$R=R_1\cup R_2$ and such that
$A(x.)$ can be expressed as a product of two terms
as follows:

\beq
A(x_{j_1}, x_{j_2}, \xvec{R})=
\alpha_1(x_{j_1}, \xvec{R_1})
\alpha_2(x_{j_2}, \xvec{R_2})
\;.
\label{eq-amp-2factors}
\eeq
As usual, let
$A(x.) = \rootp(x.)e^{\theta(x.)}$.
The previous equation implies that
$P(x.)$  can be expressed as a product of two terms
as follows:

\beq
P(x_{j_1}, x_{j_2}, \xvec{R})=
q_1(x_{j_1}, \xvec{R_1})
q_2(x_{j_2}, \xvec{R_2})
\;.
\label{eq-prob-2prod-j12}
\eeq
Summing both sides of Eq.(\ref{eq-prob-2prod-j12})
over $x_{j_2}$, over $x_{j_1}$
and over both, gives, respectively,

\beq
P(x_{j_1}, \xvec{R})=
q_1(x_{j_1}, \xvec{R_1})
\tilde{q}_2(\xvec{R_2})
\;,
\label{eq-prob-2prod-j1}
\eeq

\beq
P(x_{j_2},\xvec{R})=
\tilde{q}_1(\xvec{R_1})
q_2(x_{j_2}, \xvec{R_2})
\;,
\label{eq-prob-2prod-j2}
\eeq
and

\beq
P(\xvec{R})=
\tilde{q}_1(\xvec{R_1})
\tilde{q}_2(\xvec{R_2})
\;.
\label{eq-prob-2prod-j}
\eeq
From Eqs.(\ref{eq-prob-2prod-j12})
to (\ref{eq-prob-2prod-j}), it is clear that

\beq
P(x_{j_1}, x_{j_2} | \xvec{R})=
P(x_{j_1} | \xvec{R})
P(x_{j_2} | \xvec{R})
\;.
\label{eq-prob-indep-j12}
\eeq
Eq.(\ref{eq-amp-2factors}) implies that
$\theta(x.)$ can be expressed as a sum
of two terms as follows:

\beq
\theta(x_{j_1}, x_{j_2},\xvec{R})=
\omega_1(x_{j_1}, \xvec{R_1})+
\omega_2(x_{j_2}, \xvec{R_2})
\;.
\label{eq-theta-2sum}
\eeq
Eq.(\ref{eq-theta-2sum})
immediately yields

\bsub
\beqa
\theta(x_{j_1}, x_{j_2}|\xvec{R})&=&
\theta(x_{j_1}, x_{j_2}\xvec{R})
-
\theta(x^o_{j_1}, x^o_{j_2},\xvec{R})\\
&=&\left\{
\begin{array}{l}
[\omega_1(x_{j_1}, \xvec{R_1})+
\omega_2(x_{j_2}, \xvec{R_2})]\\
-[\omega_1(x^o_{j_1}, \xvec{R_1})+
\omega_2(x^o_{j_2}, \xvec{R_2})]
\end{array}
\right.
\;,
\eeqa
\esub

\bsub
\beqa
\theta(x_{j_1}|\xvec{R})&=&
\theta(x_{j_1}, x^o_{j_2}|\xvec{R})\\
&=&
\omega_1(x_{j_1}, \xvec{R_1})-
\omega_1(x^o_{j_1}, \xvec{R_1})
\;,
\eeqa
\esub
and

\bsub
\beqa
\theta(x_{j_2}|\xvec{R})&=&
\theta(x^o_{j_1}, x_{j_2}|\xvec{R})\\
&=&
\omega_2(x_{j_2}, \xvec{R_2})-
\omega_2(x^o_{j_2}, \xvec{R_2})
\;.
\eeqa
\esub
Thus,

\beq
\theta(x_{j_1}, x_{j_2}| \xvec{R})=
\theta(x_{j_1}| \xvec{R})
+
\theta(x_{j_2}| \xvec{R})
\;.
\label{eq-theta-indep-j12}
\eeq
Combining Eqs.(\ref{eq-prob-indep-j12}) and
(\ref{eq-theta-indep-j12}),
we get

\beq
A(x_{j_1}, x_{j_2}| \xvec{R})=
A(x_{j_1}|\xvec{R})
A(x_{j_2}|\xvec{R})
\;.
\eeq
\qed

\subsection{Going Global}

In the last section, we showed
that a probability distribution $P$
(or a probability amplitude $A$) factors according
to an UG iff it satisfies a
certain non-global, graphic I-set.
Does a similar result hold
if the non-global
 graphic I-set
 is replaced by a global graphic one?
This section will be devoted to
answering this question.

Consider any $G\in UG(\rvxdot)$.
Suppose $J,K,E\subset \zn$ are disjoint sets.
Let $I = (\rvxvec{J}\perp \rvxvec{K}|\rvxvec{E})$.
We define the function
$\tausep{}:\cali(\rvxdot)\rarrow Bool$
by the statement:
$\tausep{}(I)$ is true
iff all
paths $\gamma$ in $G$ from a node
in $\rvxvec{J}$ to a node in $\rvxvec{K}$
are blocked by $\rvxvec{E}$.
We
say ``$\gamma$ is blocked by $\rvxvec{E}$" iff
there exists a node $\rvx_i\in\gamma$ that
satisfies $i\in E$.

\begin{theo}\label{th-prob-glo-iff-loc-iff-pair}
Suppose $G\in UG(\rvxdot)$ and
$P\in pd(St_\rvxdot)$. For $\xi \in\{ glo, loc, pair\}$,
let $\Phi_{\xi}$ denote the
statement $\cali_{\xi}(G)\subset \cali(P)$.
\newline
$\Phi_{glo}\Rarrow \Phi_{loc}\Rarrow \Phi_{pair}$.
\newline
If $P\neq 0$,
$\Phi_{glo}\Leftrightarrow \Phi_{loc}\Leftrightarrow \Phi_{pair}$.
\end{theo}
\proof The proof is a special case
of the proof of the next theorem.
\qed

\begin{theo}\label{th-amp-glo-iff-loc-iff-pair}
Suppose $G\in UG(\rvxdot)$ and
$\mu\in dm(\calh_\rvxdot)$ is a
meta density matrix of the form
$\mu = \proj(\sum_{x.} A(x.)\ket{x.})$.
For $\xi \in\{ glo, loc, pair\}$,
let $\Phi_{\xi}$ denote the
statement $\cali_{\xi}(G)\subset \cali(A)$.
\newline
$\Phi_{glo}\Rarrow \Phi_{loc}\Rarrow \Phi_{pair}$.
\newline
If $A\neq 0$,
$\Phi_{glo}\Leftrightarrow \Phi_{loc}\Leftrightarrow \Phi_{pair}$.
\end{theo}
\proof

{\bf proof that $\Phi_{glo}\Rarrow \Phi_{loc}$}:
Let $I=(\rvx_j \perp
\rvxvec{\zn - \overline{ne}(j)}| \rvxvec{ne(j)})
\in \cali_{loc}(G)$.
$\tausep{}(I)$ so $\tau^A(I)$.

{\bf proof that $\Phi_{loc}\Rarrow \Phi_{pair}$}:
Suppose $j_1,j_2\in \zn$ and $j_1\not\in ne(j_2)$.
$\tau^A(\rvx_{j_1} \perp
\rvxvec{\zn - \overline{ne}(j_1)}| \rvxvec{ne(j_1)})$
and
$j_2\in \zn - \overline{ne}(j_1)$
so, using the reduction rule $2\rarrow 1'$,
we get
$\tau^A(\rvx_{j_1}\perp\rvx_{j_2}|
\rvxvec{\zn-\{j_1,j_2\}})$.

{\bf proof that $(A\neq 0, \Phi_{pair})\Rarrow \Phi_{glo}$}:
Suppose $J,K,E\subset \zn$ are disjoint sets.
Let $I=(\rvxvec{J} \perp \rvxvec{K}| \rvxvec{E})$.
Note that
$\Phi_{pair}$ is equivalent to:
$\tau^A(\rvx_{j_1}\perp\rvx_{j_2}|
\rvxvec{\zn-\{j_1,j_2\}})$ for $j_1\not\in ne(j_2)$.
What we want to prove is
$\Phi_{sep}$, which is equivalent to:
$\tausep{}(I)\Rarrow \tau^A(I)$.

If we can prove the theorem when $|J|+|K|+|E|=N$,
then the other cases will follow. This is why.
Suppose $|J|+|K|+|E|<N$ and $r\in \zn-J-K-E$.
Assume $\tausep{}(I)$.
Since $\tausep{}(I)$ is true,
either
$\tausep{}(\rvxvec{J},\rvx_r \perp \rvxvec{K}| \rvxvec{E})$
or
$\tausep{}(\rvxvec{J} \perp \rvxvec{K},\rvx_r| \rvxvec{E})$
must be true.
For if both were false, there would be
a path from a node in $\rvxvec{J}$ to a node
in $\rvxvec{K}$ that was not blocked by $\rvxvec{E}$,
contradicting $\tausep{}(I)$.
In general, all the nodes
that are not in $\rvxvec{J},\rvxvec{K}, \rvxvec{E}$,
can be put in either the $J$
side (if they are d-separated from the $K$ side)
or the $K$ side (if they are d-separated from the $J$ side).
Thus, there exist disjoint sets
$J_{fat}$ and $K_{fat}$
such that
$J_{fat}\supset J$, $K_{fat}\supset K$,
$|J_{fat}| + |K_{fat}| + |E|=N$,
and such that
$I_{fat}=
(\rvxvec{J_{fat}}\perp \rvxvec{K_{fat}}| \rvxvec{E})$
satisfies
$\tausep{}(I_{fat})$.
If we can prove that
$\tausep{}(I_{fat})\Rarrow\tau^A(I_{fat})$, then,
by virtue of the reduction rule $2\rarrow 1$,
$\tau^A(I)$ will follow.

It now remains for us to prove the theorem for
the fat case when $|J|+|K|+|E|=N$.
The proof is by induction in $|J|+|K|$.

When $|J|+|K|=2$,
$J=\{j\}$, $K=\{k\}$,
$I=(\rvx_j \perp \rvx_k| \rvxvec{E})$.
Assume $\tausep{}(I)$.
It follows that $j\not\in ne(k)$.
Hence, $I\in \cali_{pair}(G)$. Hence, $\tau^A(I)$.

Now assume
$\tausep{}(I)\Rarrow \tau^A(I)$
when $|J|+|K|\in Z_{2,\alpha}$
and try to prove it for
$|J|+|K|=\alpha+1>2$.
Either $|J|$ or $|K|$ is greater than
two, so we may assume, without loss of
generality, that $|K|>2$.
Let $k\in K$ and $K' = K-\{k\}$.
Let
$I_1 =(\rvxvec{J} \perp \rvxvec{K'}| \rvxvec{E\cup\{k\}})$,
and
$I_2 =(\rvxvec{J} \perp \rvx_k| \rvxvec{E\cup K'})$.
Assume $\tausep{}(I)$. It follows that
$\tausep{}(I_1)$ and
$\tausep{}(I_2)$ .
Furthermore,
$|J|+|K'|<\alpha+1$ and
$|J|+1<\alpha+1$ so,
by the inductive hypothesis,
$\tau^A(I_1)$ and
$\tau^A(I_2)$.
By virtue of the
combination rule $1',1'\rarrow 2$ (here
we use $A\neq 0$),
$\tau^A(I_1)$ and
$\tau^A(I_2)$ together imply $\tau^A(I)$.
\qed

\begin{theo}\label{th-mnet-prob-dsep}
(Classical d-Separation Theorem)
Suppose $G\in UG(\rvxdot)$ and
$P\in pd(St_\rvxdot)$.
If $P\neq 0$ and $P$ factors according to $G$, then
$\cali_{glo}(G)\subset \cali(P)$.
\end{theo}
\proof
Follows from Theorems
\ref{th-prob-fact-iff-local-iset}
and
\ref{th-prob-glo-iff-loc-iff-pair}.
\qed

\begin{theo}\label{th-mnet-amp-dsep}
(Quantum d-Separation Theorem)
Suppose $G\in UG(\rvxdot)$ and
$\mu\in dm(\calh_\rvxdot)$ is a
meta density matrix of the form
$\mu = \proj(\sum_{x.} A(x.)\ket{x.})$.
If $A\neq 0$ and $A$ factors according to $G$, then
$\cali_{glo}(G)\subset \cali(A)$.
\end{theo}
\proof
Follows from Theorems
\ref{th-amp-fact-iff-local-iset}
and
\ref{th-amp-glo-iff-loc-iff-pair}.
\qed

\section{d-Separation and Quantum Entanglement}

In this section, we
show that the quantum d-separation rules
for Bayesian and Markov graphs
can be used to detect
pairs $(\rvxvec{J},\rvxvec{K})$
in a graph
that are unentangled.

For $G\in DAG(\rvxdot)$, define

\beq
\cald_{glo}(G)=
\{(\rvxvec{J},\rvxvec{K}):
J,K\subset\zn{\rm\;are\;disjoint}\;;
\tausep{}(
\rvxvec{J}\perp\rvxvec{K}|\rvxvec{\zn-J-K})
\}
\;.
\eeq
For $G\in UG(\rvxdot)$, define
$\cald_{glo}(G)$ in the same way.
The function $\tausep{}$
has been defined previously.
Its definition is different for
DAGs than for UGs.

For a meta-density matrix
$\mu = \proj(\sum_{x.} A(x.)\ket{x.})$,
define

\beq
\cald(A)=
\{
(\rvxvec{J},\rvxvec{K}):
J,K\subset\zn{\rm\;are\;disjoint}\;;
E^{CMI}_\mu(\rvxvec{J}:\rvxvec{K})=0
\}
\;.
\eeq

\begin{theo}\label{th-dsep-ent}
Suppose $G\in DAG(\rvxdot)$ (ditto,
$G\in UG(\rvxdot)$)
and
$\mu\in dm(\calh_\rvxdot)$ is a
meta density matrix of the form
$\mu = \proj(\sum_{x.} A(x.)\ket{x.})$.
\newline If
$A$ factors according to $G$,
then
$\cald_{glo}(G)\subset \cald(A)$.
\end{theo}
\proof
Assume $A$ factors according to
$G\in DAG(\rvxdot)$ (ditto,
$G\in UG(\rvxdot)$). Let $J,K\subset \zn$
be disjoint sets. Let $E= \zn-J-K$.
Let $I=(\rvxvec{J} \perp \rvxvec{K}| \rvxvec{E})$.
Assume $\tausep{}(I)$.
The quantum d-separation theorem, namely
Theorem \ref{th-bnet-amp-dsep}
(ditto, Theorem \ref{th-mnet-amp-dsep}),
tells us that if $\tausep{}(I)$,
then $\tau^A(I)$.
But we know from Theorem
\ref{th-tau-a-tau-cmip-connection}
that, because $I$ is all-encompassing,
 $\tau^A(I)= \tau^{CMI'}(I)$.
It follows that
$S_{\diag_\rvxvec{E}(\mu)}
(\rvxvec{J}:\rvxvec{K}|\rvxvec{E})=0$.
This and the definition of CMI entanglement
imply that
$E^{CMI}_{\mu}
(\rvxvec{J}:\rvxvec{K})=0$.
\qed

Suppose $J\subset \zn$,
and we are given a density matrix
$\rho \in dm(\calh_{\rvxvec{J}})$ with a
{\it generalized} purification
$\mu \in dm(\calh_{\rvxdot})$.
Suppose $J_1,J_2\subset J$ are disjoint sets,
and we want to decide
whether $E^{CMI}_{\rho}(\rvxvec{J_1}:\rvxvec{J_2})$
vanishes.
Note that
to apply Theorem \ref{th-dsep-ent},
we should first replace $\mu$ by
a {\it traced} purification $\tilde{\mu}$ of $\rho$.
The reason is that we are interested in
$E^{CMI}_{\rho}(\rvxvec{J_1}:\rvxvec{J_2})$.
This quantity is not necessarily equal to
$E^{CMI}_{\mu}(\rvxvec{J_1}:\rvxvec{J_2})$,
but it is always equal to
$E^{CMI}_{\tilde{\mu}}(\rvxvec{J_1}:\rvxvec{J_2})$.

In a nutshell,
Theorem \ref{th-dsep-ent} tells us that,
if $J,K\subset\zn$ are disjoint sets, and
$\tausep{}(\rvxvec{J} \perp \rvxvec{K}|
\rvxvec{\zn-J-K})$,
then $E^{CMI}_{\mu}
(\rvxvec{J}:\rvxvec{K})=0$.
And now, some examples.
Let $k=?$ mean that we can't conclude anything
about the value of $k$.
The
Bayesian nets
\beq(\rvx)\larrow(\rva)\rarrow(\rvy)\eeq
and
\beq(\rvx)\rarrow(\rva)\rarrow(\rvy)\eeq
both have
$E^{CMI}_{\mu}
(\rvx:\rvy)=0$
because $\rva$ can be grounded
and $S_\mu(\rvx:\rvy|\rva)=0$.
On the other hand,
the Bayesian net
\beq(\rvx)\larrow
\stackrel{e\Sigma}{(\rva)}
\rarrow(\rvy)\eeq
is equivalent to
$\mu = (\rvx)\larrow
(\rvy)$, for which  $S_\mu(\rvx:\rvy)=?$,
so
$E^{CMI}_{\mu}
(\rvx:\rvy)=?$.
The Bayesian net
\beq(\rvx)\rarrow
(\rva)
\larrow(\rvy)\eeq
also has
$E^{CMI}_{\mu}
(\rvx:\rvy)=?$,
because grounding $\rva$
allows transmission of
information between $\rvx$ and $\rvy$.

A parting observation: Suppose $J,K_1,K_2\subset \zn$
are disjoint sets. Let
$D_1=(\rvxvec{J},\rvxvec{K_1})$,
$D_2=(\rvxvec{J},\rvxvec{K_2})$, and
$D=(\rvxvec{J},\rvxvec{K_1\cup K_2})$.
It's
not hard to convince oneself that
$[D\in\cald_{glo}(G)] \Leftrightarrow [D_1,D_2\in\cald_{glo}(G)]$.
By the synergism of
entanglement,
$[D\in\cald(A)]\Rarrow [D_1,D_2\in\cald(A)]$,
but
the opposite implication does not appear to be
true. If we define a perfect graph as
one for which
$\cald_{glo}(G)=\cald(A)$,
then it appears that
no all graphs are perfect.

\newpage
\appendix
\section{Appendix: Mobius Inversion}
\label{app-mobius}

In this appendix, we prove
the Mobius Inversion Theorem.

Some preliminary observations
will facilitate our proof.

\begin{figure}[h]
    \begin{center}
    \epsfig{file=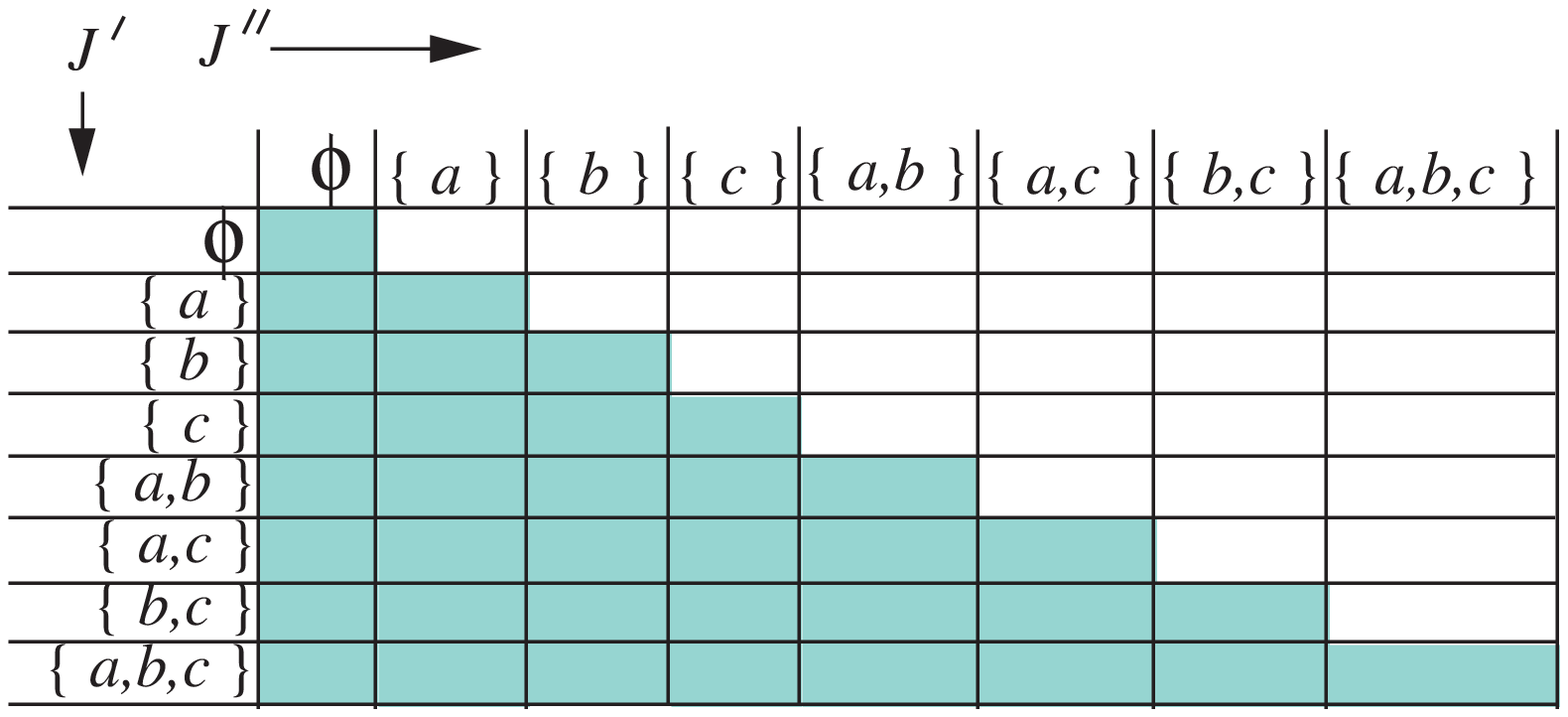, height=2in}
    \caption{Table with rows and columns labelled
    by all the subsets
    of the set $J=\{a,b,c\}$. We want
    to sum over the shaded entries
    of this table.
    }
    \label{fig-powset-table}
    \end{center}
\end{figure}

For any finite set $J$,
consider a table  of arbitrary
complex numbers, where the
rows and columns of the table are both
labelled by the elements of
$2^J$. Suppose we want to
sum over the elements of
the table  that are
below its main diagonal.
Fig.\ref{fig-powset-table} illustrates the table
for $J=\{a,b,c\}$.
The shaded entries of Fig.\ref{fig-powset-table}
are the entries we want to sum over.
Two simple methods for carrying out such a sum
are:(1)
sum first over rows
and then over columns,
(2) sum first over columns and then over rows.
Of course, whether we use
method (1) or (2), the
final value of the sum will not change.
This simple observation, that
the final value of the sum does not depend
on the order of summation, can be
stated more formally as

\beq
\sum_{J':J'\subset J}\;\;
\sum_{J'':J''\subset J'}
=
\sum_{J'':J''\subset J}\;\;
\sum_{J':J''\subset J'\subset J}
\;.
\eeq
By
$\sum_{J':J'\subset J}$ we mean the sum of
all subsets $J'$ of $J$, including the empty
set $\emptyset$ and $J$. Note that we
use $J',J''$ (i.e., $J$ with one or more
primes) to denote subsets of $J$.

\begin{figure}[h]
    \begin{center}
    \epsfig{file=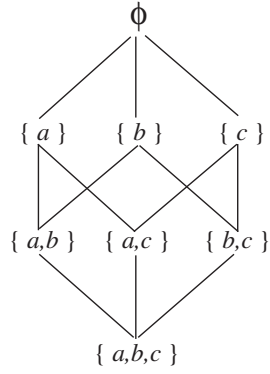, height=2in}
    \caption{All subsets of $\{a,b,c\}$,
    arranged on a lattice.}
    \label{fig-powset-lattice}
    \end{center}
\end{figure}

Another simple observation is that

\beq
\sum_{D': D'\subset D}
(-1)^{|D'|} =
\delta(D, \emptyset)
\;.
\label{eq-mobius-delta-fun}
\eeq
For example, suppose
$D=\{a,b,c\}$. Fig.\ref{fig-powset-lattice}
lists all the subsets $D'$ of $D$.
It associates each distinct $D'$
with a different node of a lattice.
(Subsets with the same number of
elements are in the same horizontal level.
Subsets in lower horizontal levels
have more elements.
Links connect subsets that differ only by one
element.)
If we sum $(-1)^{D'}$
over all the nodes of the
lattice of Fig.\ref{fig-powset-lattice}, we
get zero, since the number of
even-order subsets
equals the number of odd-order subsets.
This is true for any set $D$
except for the empty set, which
has only a single even-order subset, itself.
Eq.(\ref{eq-mobius-delta-fun}) yields

\bsub
\beqa
\sum_{J':J''\subset J'\subset J}
(-1)^{|J'-J''|}&=&
\sum_{\Delta J' :
\emptyset
\subset \Delta J'
\subset \Delta J=J-J''}
(-1)^{|\Delta J'|}\\
&=&\delta(J,J'')
\;.
\eeqa
\esub
If $J''\subset J' \subset J$, then
$|J-J'| + |J'-J''| + |J''| = |J|$
so

\bsub
\beqa
\sum_{J':J''\subset J' \subset J}
(-1)^{|J-J'|}&=& (-1)^{|J|-|J''|}
\sum_{J':J''\subset J'\subset J}
(-1)^{-|J'-J''|}\\
&=&\delta(J,J'')
\;.
\eeqa
\esub

\begin{theo}
For any set $J$,
and any functions $f,g:2^J\rarrow \CC$,

\bsub
\beq
g(J) = \sum_{J': J'\subset J}(-1)^{|J-J'|} f(J')
\;
\eeq
if and only if

\beq
f(J) = \sum_{J': J'\subset J} g(J')
\;.
\eeq
\esub
\end{theo}
\proof

($\Rarrow$)
\bsub
\beqa
\sum_{J': J'\subset J} g(J') &=&
\sum_{J': J'\subset J}\;\;
\sum_{J'': J''\subset J'}(-1)^{|J'-J''|} f(J'')\\
&=&
\sum_{J'':J''\subset J}\;\;
\sum_{J':J''\subset J'\subset J}
(-1)^{|J'-J''|} f(J'')\\
&=&
f(J)
\;
\eeqa
\esub

($\Larrow$)
\bsub
\beqa
\sum_{J': J'\subset J}(-1)^{|J-J'|} f(J')&=&
\sum_{J': J'\subset J}(-1)^{|J-J'|}
\sum_{J'': J''\subset J'} g(J'')\\
&=&
\sum_{J'': J''\subset J}\;\;
\sum_{J':J''\subset J' \subset J}
(-1)^{|J-J'|}g(J'')\\
&=&
g(J)
\;
\eeqa
\esub
\qed


\begin{thebibliography}{99}

\bibitem{Kol}
Daphne Koller and Nir Friedman,
{\it Bayesian Networks and Beyond},
to be published.

\bibitem{Lau}
S. L. Lauritzen,{\it Graphical Models}
(Clarendon Press, Oxford, 1996).

\bibitem{Sew}http://en.wikipedia.org/wiki/Sewall\_Wright

\bibitem{Pea-wiki}http://en.wikipedia.org/wiki/Judea\_Pearl

\bibitem{Pea1}Judea Pearl,
{\it Probabilistic Reasoning in Intelligent Systems}
(Morgan-Kaufmann, 1988).

\bibitem{Pea2}Judea Pearl,
{\it Causality: Models, Reasoning, and Inference}
(Cambridge University Press, 2000).

\bibitem{hist}
www.andrew.cmu.edu/user/scheines/tutor/d-sep.html

\bibitem{Tucci-bnet1}
Robert R. Tucci,
``Quantum Bayesian Nets",
quant-ph/9909039

\bibitem{Tucci-meta-rho1}
Robert R. Tucci,
``Quantum Information Theory - A Quantum Bayesian Net Perspective",
quant-ph/9909039

\bibitem{Tucci-bnet-qcomp}
Robert R. Tucci,
``Quantum Computer as a Probabilistic Inference Engine",
quant-ph/0004028

\bibitem{artiste}http://www.ar-tiste.com

\bibitem{Kat}Kathryn Blackmond Laskey,
 http://ite.gmu.edu/\~{}klaskey,
 ``Quantum Physical Symbol Systems", Journal of Logic, Language and Information, 2005;
 ``Quantum Causal Networks", to be published

\bibitem{wiki-cmi-ent}
http://en.wikipedia.org/wiki/Squashed\_entanglement

\bibitem{Cover}T.M. Cover, J.A. Thomas,
{\it Elements of Information Theory} (1991, John Wiley).


\bibitem{Tucci-separability}
Robert R. Tucci,
``Separability of Density Matrices and Conditional Information Transmission",
quant-ph/0005119

\bibitem{Petz}
P. Hayden, R. Jozsa, D. Petz, A. Winter,
``Structure of states which satisfy strong subadditivity of quantum entropy with equality",
quant-ph/0304007

\end{thebibliography}
\end{document}